\documentclass[final,3p,times]{elsarticle}
\usepackage{epsfig,lineno}

\usepackage{amssymb}

\journal{Astroparticle Physics}

\begin{document}

\begin{frontmatter}



\title{Mapping the Most Energetic Cosmic Rays}


\author{A.M. Hillas}

\address{School of Physics and Astronomy, University of Leeds, Leeds LS2 9JT,  UK}

\begin{abstract}
The Pierre Auger Collaboration has shown that the cosmic rays detected to August 2007, with estimated
energies above 57 EeV, were mostly very close to the direction of a catalogued AGN within 
$\sim 75$ Mpc.  
The closeness of the sources to us, and their association with the locality of moderate Seyfert 
galaxies rather than the most striking radio galaxies, were surprising, leading some 
authors to question the reality of the apparent associations.  
Here, three further techniques for examining the correlation of cosmic ray arrival directions 
with directions of AGNs are introduced to confirm and extend this correlation.  
These include the uniform-exposure polar plot to examine large-scale associations, and 
a sensitive ``right ascension resonance'' test to show the rapidity of the decoherence when 
the two patterns are displaced.  
The latter test avoids the choice of a correlation window radius, which makes it possible
to see an AGN correlation in other data at a lower energy, and to seek it in HiRes data.
On the basis of the closeness of correspondence of the cosmic ray and AGN maps, and the equally
significant correspondence with directions of extended radio galaxies
listed by Nagar and Matulich but not used in 
the Auger group's investigation, it is argued that the association with 
these two sets of objects is by no means accidental, although the efficacy of the
57 EeV ``cut'' in selecting this revelatory sample may have been accidental.

The arrival directions of these cosmic rays (nominally close to 75 EeV) can be well described if most of the sources 
are in or around rather typical Seyfert galaxies, in clusters typically at 
$\sim 50$ Mpc, with the cosmic rays being 
scattered by $3-4^\circ$ on their way to us.  
Because of close clustering of AGNs, it cannot usually be ascertained which object
within 2--3 Mpc is the actual source, but more than a third of the cosmic rays 
appear to come from FRI or similar radio galaxies in the clusters.
It thus seems likely that these and also weaker jet-forming Seyfert galaxies
are indeed causing acceleration to $10^{20}$ eV. 
Neither the brightest (nearby) radio galaxies nor the Virgo cluster dominate the cosmic-ray sky as had 
been expected, and Cen A is probably one of the currently inactive $10^{20}$ eV accelerators, 
as much more distant FRI galaxies play such a large role.  
The deflections cannot be much more than $3-4^\circ$ without destroying the coherence.  
Intergalactic magnetic fields $\sim$1 nG could be responsible, but in one part of the sky a 
displaced ``resonance'' suggests a possible $4^\circ$ deflection by the $B_z$ component of
our local Galactic magnetic field.  
A source region limited to $<$120 Mpc for the Auger cosmic rays is supported.  
This is compatible with a GZK survival horizon but only if (a) the sudden fall in the energy 
spectrum is not simply a GZK effect but essentially reflects the energy cut-off in the accelerators,
and (b) the Auger energies are underestimated by $\sim 25\%$. 

\end{abstract}


\end{frontmatter}
\linenumbers

\section{Introduction to the new observations of arrival directions of 
the most energetic cosmic rays}
\label{introsec}



In 2007 the Pierre Auger Collaboration \cite{auger07} published the directions of arrival 
of 27 cosmic-ray particles which they estimated to have energies above 
$5.7 \times 10^{19}$ eV (57 EeV), and near most of these directions (within $3.2^\circ$) 
there was an active galactic nucleus (AGN) listed in the 12th V\'{e}ron-Cetty  V\'{e}ron 
catalogue (abbreviated below as VCV) \cite{vvcat} within a distance of about 75 Mpc of 
Earth (if $H_o=72\ \mathrm{km\ s^{-1} Mpc^{-1}}$, as assumed below).  The probability of such a 
close association appeared to be very small if the cosmic rays arrived 
from near-random directions as only 21\% of cosmic rays would be expected to match so 
closely in this case. By contrast, many air shower experiments had previously sought evidence for 
the sources of cosmic rays above $10^{17}$ eV, but had failed to find any convincing departures from isotropy,
supporting the belief that deflections of the particles in irregular magnetic fields had scrambled 
their directions of motion.  
Above about 60-80 EeV, though (depending on the maximum energy of particles leaving the sources), 
few protons or light nuclei would reach us from beyond very few hundred Mpc because of energy losses 
in the cosmic microwave background radiation (the GZK effect), and so the 
particles from far off, that would have been highly deflected en route, should be absent.
The Auger experiment in Argentina is unique in its huge collecting aperture, its very uniform 
sensitivity to cosmic rays above $10^{19}$ eV (within $60^\circ$ of the zenith), 
well-measured energies in this context, and, apparently, an inspired unexpected choice of the 
VCV catalogue to check for significant patterns in the arrival directions.  
The data set will be greatly expanded in the next few years, but it already shows that, firstly, 
the most spectacular double-lobe radio galaxies that dominate the radio sky are not the 
principal suppliers of these extreme cosmic rays, secondly, the expected beacon of the local skies, 
the Virgo cluster, is also relatively unimportant, and in fact most of these extreme cosmic rays 
appear to originate in less remarkable Seyfert galaxies or else in sources that cluster in their 
vicinities (say 2-3 Mpc).  Looking outside the VCV catalogue of optically-detected AGNs used in the 
Auger report, it is found additionally that a considerable minority of these cosmic rays correlate about 
as closely, and perhaps even more significantly, with extended radio galaxies within the same distance 
range, although somewhat further away than those which dominate the radio sky:
a set of sources brought to our attention by Nagar and Matulich \cite{nagmat}.  
Very fortunately, it seems that, at these energies, deflections of the particles over 100 Mpc are not more 
than a few degrees.
However, it will be proposed that this description applies only to the proton component of
cosmic rays, whereas there is evidence that above 40 EeV most particles are highly
charged; so a detector's bias in favour of protons may strongly influence the pattern that it sees.

It is a novel feature for charged-particle astronomy to reveal a rich pattern of point sources, 
and these surprises have led to challenges.  It is the first main object of this paper to provide further 
methods of analysis that greatly strengthen the evidence that these details of the arrival patterns are real, 
and then to examine the most probable implications regarding 
the origin, nature and energies of the particles, that can be straightforwardly tested 
with more data.

The unexpectedness of these reported correlations led commentators to express doubts of many kinds.  
First, regarding the methodology, it has been suspected that the apparent statistical significance of 
the result has been artificially generated by the use of the first half of the Auger observational data set 
to vary the selection ``cuts'' in energy threshold (57 EeV), maximum redshift of catalogued AGNs 
to be used in the correlation ($< 0.018$), and the correlation window size ($3.2^\circ$): all chosen to maximize 
the significance of the departure from isotropy in this first subset.  
Several early discussions of the arrival directions have put aside the apparent close correlation with
weak AGNs, and considered large magnetic displacements from the directions of expected 
strong sources \cite{ biermann3,biermann4,gorbunov}.  
The use of the VCV catalogue based on optical appearance of galaxies, rather than one with more emphasis 
on high jet luminosity indicated by strong radio or X-ray emission, has been questioned  \cite{swift,moska}.  
Different proportions of AGNs will be included in the catalogue in different regions of space. 
A deduction that most of the sources are less than 100 Mpc away has been remarked on as 
strange \cite{stan}.  The northern-hemisphere HiRes experiment found no association of the most energetic 
cosmic rays with AGNs \cite{hires}.  The very weak level of activity of some of the AGNs picked out in the
study has been remarked on as making their identification as cosmic-ray sources implausible.  
(It should be pointed out that the Auger authors did not claim to discriminate between the AGNs or other 
objects that reside in the same localities as the AGNs.  
Their published study was intended to demonstrate anisotropy rather than to explain it at this stage.)

Whilst the aim of the research is to understand how particles are accelerated to these extreme 
energies \cite{biermann3,biermann4,stan}, the emphasis in this paper is on extracting as clearly 
as possible the simplest pointers to sources that are present in the cosmic-ray directional data, 
largely avoiding scrutiny of particular accelerators.  
When three times the presently published data set is available, it should be possible to pay attention 
to individual sources.  
So it is the purpose in this paper to tackle many of these doubts, and to show that the evidence can be 
extended.  The clumpy pattern on the sky of these AGNs can be demonstrated by a simple extension of the 
Auger ``window'' counts, and this clumpy distributions of cosmic rays and AGNs on the sky can be better 
visualised and quantified by an alternative form of mapping, using a uniform-exposure polar plot.
A simple model of apparent origins of cosmic rays scattered around VCV AGN locations describes also the 
overall pattern of the cosmic rays.  
(The clumping also implies that not all AGNs near the cosmic-ray line of sight are necessarily the relevant 
sources, so several objects can have been misidentified as sources.)  
A ``right ascension resonance'' test is introduced as a striking and sensitive demonstration of the 
association with AGN localities, using which such associations can later be sought using a different 
list of potential sources, in data of other experiments, and potentially at other energies, and a hint of magnetic 
deflection appears.

The work is laid out as follows.  
The direct evidence of the density and clustering of AGNs near cosmic-ray directions is examined in 
section \ref{clustersec}, showing that the particles do not come precisely from the AGNs, but are 
distributed within $\sim 3-4^\circ$.  
In section \ref{ueppsec}, the use of the uniform-exposure polar plot is introduced, to show where 
these clusters are on the sky, and demonstrating first the anisotropy of the cosmic-ray distribution without 
regard to AGNs, and then that its large-scale features do match those of the AGNs in the VCV catalogue 
(which has a ``useful'' distance bias), apart from a possible anomaly in the direction of the Virgo cluster, 
which will be re-visited later.  
Section \ref{rare4sec} introduces a ``right ascension resonance'' as an alternative to the Auger method of 
detecting a correlation of cosmic ray directions with AGNs, but not requiring the choice of a 
particular ``window radius'' for counting nearby AGNs, and avoiding the large statistical penalty which 
the choice of this radius involves. 
Using this, an AGN association can be sought in the results of two experiments
viewing the northern hemisphere. 
In section \ref{radgal5sec}, the correlation of cosmic rays with an alternative set of AGNs, 
``extended radio galaxies'' \cite{nagmat},  will be considered. 
In section \ref{mag6sec} the resonance technique is used to seek evidence of magnetic deflections where 
particles arrive after following a long path just above the disc of our galaxy.  
Section \ref{dist7sec} explores the extent in redshift of the AGN correlation, finding rough agreement 
with the Auger group's conclusion that most of the particles reaching us at these energies have travelled less than 100 Mpc.  
In section \ref{implic8sec}, it is argued that this implies an underestimation of particle energies, 
and also implies that the reported steep fall in the cosmic-ray spectrum is not simply a ``GZK spectrum cutoff'', 
but largely a downturn of the energy spectrum of the accelerators, which imposes a ``GZK distance horizon''; 
but the details are sensitive to probable energy measurement errors.  
Some principal conclusions are summarized in section \ref{summ9sec}.

Appendix \ref{Avvcsec} introduces some relevant shortcomings of the VCV catalogue of optically-detected  AGNs, 
which limit the range of source distance for which this catalogue can be used.  
Figure \ref{vcvcharac} in the appendix shows the number of objects visible from the southern hemisphere 
that are listed per Mpc of distance and in different ranges of absolute magnitude, which indicates that 
the completeness of the catalogue for fainter objects (some of which may nevertheless be particle accelerators) 
is continually decreasing with increasing distance, and that beyond about 75 Mpc the gaps may be serious, 
so that it will be more difficult to use the VCV catalogue to explore further into space 
(as remarked in \cite{auger07}).  
If the number of listed AGNs per square degree had been much greater, however, the probability of associations 
within $3.2^\circ$ with any arbitrary direction would have been large, nullifying the signal.  
A sparser but uniform catalogue would be better for such an exploration, and it will be seen that a catalogue 
of FRI radio galaxies may serve well here.  
Indeed following up the suggestion of Nagar and Matulich \cite{nagmat}, the correlation with such objects 
already adds substantially to the significance of these associations.

\section{ Clustering of the AGNs: the cosmic ray source is often not the closest AGN 
  \label{clustersec}}

The evidence that these most energetic cosmic rays originate in the locality of AGNs was that, 
excluding 6 cosmic rays arriving within $12^\circ$ of the galactic plane, where most AGNs would be obscured 
by dust, and where cosmic rays might be more strongly deflected by galactic magnetic fields, only 2 of the 
remaining 21 cosmic ray directions did not have an AGN (redshift $<$0.018) within an angular radius of $3.2^\circ$, whereas 
if one places such $3.2^\circ$ windows at arbitrary positions on the sky, governed only by the total exposure 
of the detector array to each region (but avoiding the vicinity of the galactic plane), one finds only 24\% 
of them contain AGNs.\footnote{Note that 
the proportion 21\% quoted by the Auger authors applies if the galactic obscuration zone is   
not excluded from the counts.}
(The average \emph {number} of AGNs per window is 0.36.)  

Though the statistics seem, and are, significant, such figures can always be enhanced by the selective choice 
of examples to count; and the cuts in the data --- the choice of 57 EeV as the energy threshold, of a redshift 
range below 0.018 and a window radius of $3.2^\circ$ in which to count AGNs --- were selected only when half of the 
cosmic rays had been recorded (14 of the 27 events), in such a way as to maximize the significance of the 
departure from randomness, these cuts being fixed thereafter, in order to confirm the anisotropy --- 
though actually slightly modified later (3.1 became 3.2, 56 became 57).  
There is no physical reason for expecting $3.2^\circ$ to be appropriate: this value may serve mainly to 
minimize the chance of random unassociated AGNs being counted.  
This optimization of the cuts using half the data must have enhanced the apparent association of AGNs: 
in the first half all 10/10 of the cosmic rays outside the galactic zone had an AGN in their window: 
in the last half the proportion was 9/11.  
The bias is actually not huge, but it is important to investigate it.
Also, the next paragraph will supply 
support for the supposition that the galactic belt should be excluded.

The low average density of 0.36 AGNs (in this catalogue and in this redshift range) per $3.2^\circ$ window 
might be thought to indicate that when an AGN is seen in the window it must probably be the source of the 
cosmic ray particle, but this is not so, firstly because the AGNs are strongly clustered.  
The average number of AGNs per window drawn around the 21 selected cosmic rays (``window 1'', below) is 1.1, 
but the average numbers in the 3 surrounding annuli 
$3.2^\circ -4.53^\circ, 4.53^\circ -5.54^\circ$ and $5.54^\circ -6.40^\circ$ (``windows 2 to 4'', below), 
all of equal solid angle, are given in line (A) of the table \ref{tab-1}.


\begin{table}[!h]
\begin{center}
\begin{tabular}[t]{rlrrrr}
   \hline
   & window number & 1 & 2 & 3 & 4\\
   \hline
(A) & $\langle N_{\rm AGN}\rangle$ in windows around CR & 1.1 & 1.1 & 1.0 & 0.7 \\
(B) & $\langle N_{\rm otherAGN}\rangle$ around selected AGN & 1.2 & 0.9 & 0.9 & 0.8\\
   \hline
\end{tabular}
\end{center}
\caption{Numbers of AGNs in concentric annuli.}
\label{tab-1}
\end{table}

Line (A) shows that there is typically a high local density of AGNs which extends for several more degrees.  
To find the typical density of AGNs in any region inhabited by AGNs, one can draw windows and 
surrounding annuli around any AGN, and line (B) of the table shows the average number of other AGNs 
found in these windows (the average being weighted by the array exposure at that declination, 
and avoiding the zone of low galactic latitude).  
The AGN densities around cosmic ray directions are little different: clearly an ``other AGN'', 
unrelated to any that might be the cosmic ray source, is very likely to be found 
in the central window; so the AGN seen in the standard $3.2^\circ$ window used by Auger 
may often be a neighbour unrelated to the cosmic ray.  
Indeed, if the cosmic rays always pointed back very close to a source AGN (always well 
within $3.2^\circ$), one would expect the numbers in line (A) to be like those in 
line (B) except for an additional 1 in the central window (because we are now counting 
the ``central'' AGN in addition to the ``other'' AGNs that were counted in line (B)), 
and the count in windows 1 and 2 of line (A) would be expected to differ by about 1.0, 
which they do not.  
A precaution is needed before carrying this arithmetic further, because the counts 
in the central window (window 1 in the above table) will be biased 
in the first half of the data sample, which was used to optimize the data cuts.  
(A sign of this selection bias in the first half of the data is that the average 
number of AGNs per window is 1.1 but there are no zeros --- a quite un-poissonian 
distribution!)  
Using the second half of the data, with only 11 showers outside the zone of low 
galactic latitude, the numbers of AGNs seen, this time in successive ranges of $2.5^\circ$ 
from the cosmic ray direction, have been counted, and are shown below in the line 
``OBS'', their numbers expressed as density per $9.80 \times 10^{-3}\ \mathrm{steradian}$, 
this being the area of the standard $3.2^\circ$ window.  
The excess number of counts at the closest distances is sensitive to how far cosmic rays 
are spread around a ``source'' AGN (either by magnetic scattering or by the size of a 
cloud of supposed accelerators such as magnetars or hypernovae).  
One can simulate the numbers of AGNs that would be found at different distances from 
a cosmic-ray direction if they are typically spread with r.m.s. distance $3.2^\circ , 
4^\circ\ \mathrm{or}\ 5^\circ$, for example, around AGNs chosen randomly from the catalogue, 
but then accepted with a probability matching the observational exposure to that 
declination.  
(One can only take a very simple approach in such a small sample, and the question 
of whether an AGN closer to us should have been given a higher weight will be taken up later.)  
The results are shown in table \ref{tab-2}
as ``SIM $3.2^\circ$ spread'', ``SIM $4.0^\circ$'' and 
``SIM $5.0^\circ$''.  
Of several examples calculated, the  r.m.s. spread of $4^\circ$ gave the nearest match 
to the observations, though the scanty statistics do not offer much precision
(but a spread in accordance with this will be estimated in other ways later). 

\begin{table}[!h]
\footnotesize
\begin{center}
\begin{tabular}[t]{lccccccccccccc}
\hline
 angular distance ($^\circ$) &   0  & &2.5& &5& &7.5& &10& &12.5& &15\\
\hline
 \small
OBS   & &1.49& &0.89& &0.78& &0.60& &0.53& &0.46& \\  
SIM $3.2^\circ$ spread& &1.75& &1.13& &0.79& &0.65& &0.58& &0.54& \\
SIM $4.0^\circ$ spread& &1.48& &1.11& &0.82& &0.66& &0.58& &0.54& \\
SIM $5.0^\circ$ spread& &1.24& &1.05& &0.83& &0.68& &0.60& &0.54& \\
\hline
\end{tabular}
\end{center}
\normalsize
\caption{Surface density of AGNs near cosmic ray directions.}
\label{tab-2}
\end{table}

The observations thus suggest that cosmic-ray arrival directions have an r.m.s. spread 
of about $4^\circ$ 
around typical AGNs, although this cannot be a universal constant as the AGNs will of course be at different distances.  
The AGN density remains considerably above the average density of 0.36 for $\sim 10^\circ$, indicating the 
typical extent of AGN clusters.  

Thus the cosmic rays may suffer magnetic scattering of $4^\circ$ while travelling to the observer from an AGN source.  
Since this is somewhat greater than the $3.2^\circ$ radius of the Auger selection window, one should not place 
too much emphasis on the type of AGN found closest to the cosmic ray or even in the window.  
The true source may well lie outside the window; and \emph {a prime reason why some AGN is almost always seen in the 
window is because they are so crowded in the region around the source}.  
The appearance of particularly feeble objects in the list need not cause surprise. 
Thus Zaw, Farrar and Greene \cite{farrar} noted that whilst most objects picked out by these windows were 
AGNs with bolometric luminosities above $0.5\times 10^{43}$ erg s$^{-1}$, two other neighbouring objects had been 
misidentified as AGNs.   
It is found that a very strong association with cosmic rays still persists if only radio-detected AGNs are 
retained in the catalogue (about 75\% of all those listed).  
One probably has a mixture of galaxy types close by in typical clusters, and one can deduce only that the 
sources lie in or near AGNs of a type that is very common in such clusters --- say comprising $1/3$ or more of 
the population.  
But the evidence so far does not prove that AGNs are sources.  
Since unusual activity in galactic nuclei, and growth of supermassive black holes, is believed to arise from 
collisions between galaxies in clusters, it is also possible that the AGNs are not the sources but are serving 
as markers of regions where other types of activity related to star formation are also concentrated.  
The role of AGNs as markers of concentrations of general galactic activity has been taken up by 
Ghisellini et al. \cite{ghis}, who showed that clusters of galaxies did indeed correlate with Auger cosmic ray 
directions.

Much of the following discussion will avoid the choice of a ``window'' of a particular radius, but before moving 
on it is worth mentioning that, of the windows around the 6 cosmic rays in the galactic exclusion zone, 
somewhat crudely set at a uniform belt within $12^\circ$ of the galactic plane, only 1 contains a VCV AGN and the 
first two surrounding annuli contain only 1 AGN in all (12 annuli).  
These do appear to be regions almost devoid of visible AGNs, supporting their designation as obscured regions 
where we do not know the positions of AGNs, as this catalogue is based on optical surveys, 
so tests of association cannot be performed (though the bright radio source Cen B does shine through the dust
and coincides with a cosmic ray).

Where are these AGN clusters?  
The large scale features of the cosmic ray directions and associated AGNs will be examined in the next section.

\section{Mapping the sky on a uniform-exposure polar plot
  \label{ueppsec}}

The Aitoff sky projection that is commonly used to show how cosmic-ray arrival directions relate to large-scale 
astronomical features can illustrate the relationship between clear-cut structures, but is less suited to 
these irregular and clumpy patterns.  
Not only does the varying efficiency of detecting cosmic rays in different parts of the diagram disguise real 
variations in intensity, but adjacent regions are cut apart, and important great circles such as the galactic 
plane twist sharply, making it hard to see whether the cosmic ray patterns relate to these planes. 
The pattern of cosmic ray arrival directions is revealed much better in a uniform-exposure polar plot, 
described below, if one is content to display one hemisphere, or slightly more, rather than the whole sky.

Although I first used this plot in the 1970s in correspondence about the Yakutsk observations, it was not fruitful 
then, as no patterns emerged, and I regret not having published it, although it has been used a few times in 
publications by colleagues (e.g. \cite{uep1}).   
Only now, with many particles detectable above the GZK barrier, is a study of tens, rather than thousands of 
arrival directions of great interest again.

\begin{figure}[h!]
\begin{center}
\epsfig{file=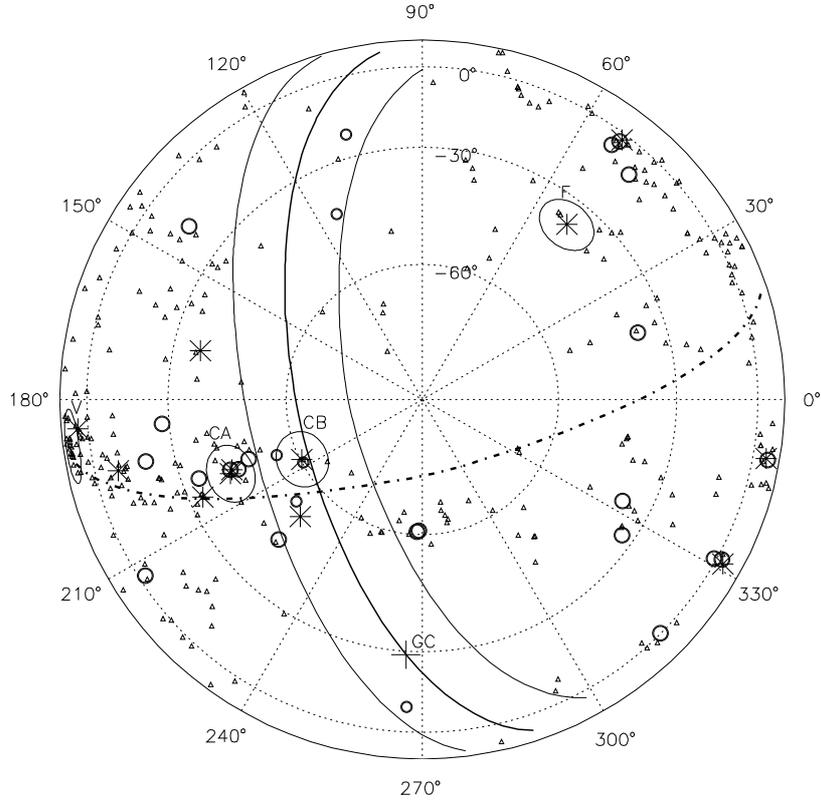,width=11.0cm}
\caption{  
The 27 arrival directions of particles above $5.7\times 10^{19}$ eV, published by 
Abraham et al. \cite{auger07} on a uniform-exposure polar plot.
The 6 particles within $12^\circ$ of the galactic plane are shown as smaller circles. 
NB:  near R.A. $268^\circ$, dec. $-61^\circ$ there are two cosmic rays extremely close together.  
The galactic plane and circles of galactic latitude $\pm 12^\circ$ are shown as full lines 
(GC marking the galactic centre) and the supergalactic plane as a dot-dash curve.  
Declination circles at $30^\circ$ intervals, and radial right ascension lines, are shown dotted.  
Circles of $6^\circ$ radius drawn around some of the strongest radio galaxies -- V,Virgo A (M87), 
CA, Cen A, CB, Cen B and F, Fornax A -- illustrate the compression of the radial scale towards the edge.
(A $6^\circ$ circle represents the angular distance from a source within which 
(at $\sim 40$  Mpc) most of these 
cosmic rays arrive according to the analyses discussed here.)  
Virgo A is at the centre of the Virgo cluster of galaxies.  
Small triangles mark the positions of 280 VCV AGNs S of declination $+18^\circ$, having redshifts $<0.018$.  
Eight-armed crosses (stars) mark the extended radio galaxies listed 
in \cite{nagmat}.}
\label{uepp1}
\end{center}
\end{figure}

Figure \ref{uepp1} shows such a plot, for the 27 Auger cosmic rays.  
The plots are always centred on a pole --- the south celestial pole in this case.  
If $\delta$ is the declination of a point in the sky, and  $p$  its polar distance, 
( $p = 90^\circ - \delta$  for a north polar plot or  $p = 90^\circ + \delta$  for a southern plot), 
the radial coordinate, $R$, in the plot, representing polar distance, is non-linear, being adjusted according 
to the total exposure of the observatory to each band of declination, so that the area of any band of $p$ is 
proportional to the number of particles that would be detected in that band if the particles arrived at 
the earth isotropically.  
Particles detected from an isotropic flux would then yield a plot with a uniform density of points per unit area, 
apart from the usual accidental statistical fluctuations.  

If, for an isotropic flux, one detects  $n_p(p)$  particles at polar distances less than $p$, out of a total 
of $n$ detected particles,
\begin{equation}  R = \sqrt{ (n_p /n)} \label{requix}
\end{equation}
(The diagram has unit radius.)  
The azimuthal angle  $\phi = \alpha$ , the right ascension.  
A comment is appropriate here for the armchair astronomer.  
With this choice of $\alpha$
(increasing anticlockwise),  this diagram looks very much like a true spherical sky globe seen from within in the case 
of the southern sky (and so represents the sky much as it really looks): but a northern hemisphere plot would 
look much like a sky globe viewed from the outside (giving a left-to-right mirror image of the constellations).  
In the north, this physicist's convention for plotting right ascension can be disconcerting to those astronomers 
who actually look at the sky, but will have the advantage of allowing the viewer to follow features such as 
the Virgo supercluster continuously from one hemisphere to the other.

If the observatory can record cosmic rays arriving at any point on a well-defined ground area, $A$, independent 
of their zenith angle of arrival, $\theta$, out to a maximum zenith angle $\theta_{max}$ --- as is the case 
for the Pierre Auger Observatory at energies above $10^{19}$ eV, where $\theta_{max} = 60^\circ$ --- the 
detection efficiency at each declination or polar distance is well defined and the radial scale $R(p)$ in 
equation \ref{requix} can be derived from first principles.
For the Auger case, with $\theta_{max} = 60 ^\circ$ and
latitude $-35.2^\circ$, the radius $R$ is tabulated in appendix \ref{A1sec}, table \ref{tab-4};
alternatively an approximate formula can be used, such as equation (\ref{rapprox2}) in the appendix,
which has a maximum error of 0.006.  
Polar angles appreciably beyond $114^\circ$ are unobservable --- and off the diagram.  
In other cases, one would generally use data on the relative numbers $n_p/n$ of detected particles in some 
well-defined lower energy band, close enough to 57 EeV (or whatever) for the detecting conditions to be 
virtually the same, but low enough for the particles to be approximately isotropic and much more numerous.  
Substitution of these numbers in equation \ref{requix} gives the radial scale $R(p)$ for the plot.

\begin{figure}[h!]
\begin{center}
\epsfig{file=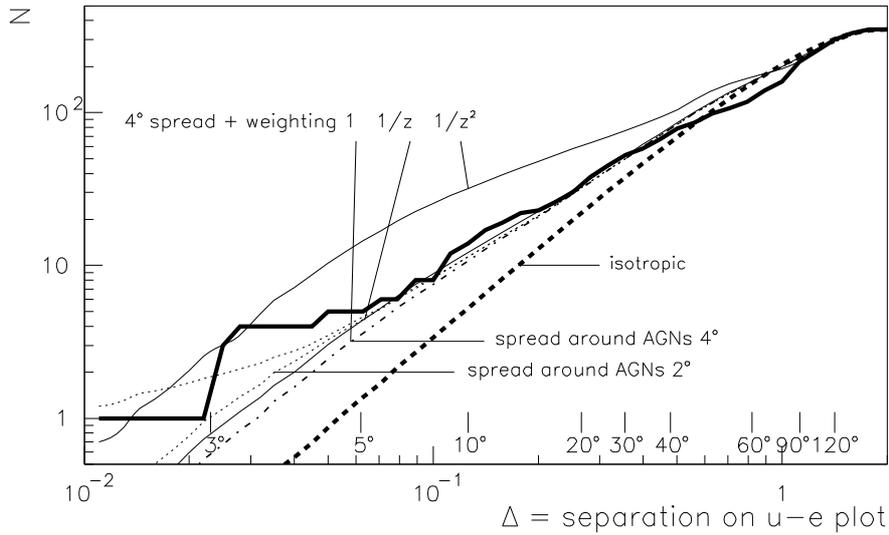,width=12.0cm}
\caption{Cumulative number N of inter-point distances $\Delta$ on uniform-exposure plot in figure \ref{uepp1} 
for the 27 Auger cosmic-ray directions.  
The average separations in degrees on the sky, to which these plot-separations $\Delta$ refer, are shown 
just above the horizontal axis. 
Thick line: the observations (fig \ref{uepp1}).  Thick dashed line (``isotropic''), shows the average
$N(\Delta)$  for many sets of 27 points placed randomly on the plot, simulating cosmic rays approaching the earth isotropically.  
Dot-dash line: average $N(\Delta)$  for many sets of 27 directions of AGNs
picked from the VCV catalogue (redshifts z$<$\ 0.018) and given a $4^\circ$ rms scatter
(with refinements described in the text).  
Changing the scatter to $2^\circ$ or $0.5^\circ$ produces the variants shown by dotted lines.
If a distance weighting is applied when picking AGNs, a $1/z^2$ weighting 
(appropriate if catalogue unbiased) moves the ($4^\circ$) curve to the upper thin full line; the preferred $1/z$ weighting 
(see text) gives the full-line curve little different from the unweighted (dot-dash) version.}
\label{kolmos2}
\end{center}
\end{figure}

Even without considering a connection with AGNs, the 27 Auger showers shown in figure \ref{uepp1} appear to be 
distributed non-randomly, but the eye and brain will usually imagine an interesting relationship in a pattern 
of dots (``canals on Mars'').  
To characterize the clumpiness of the pattern of points, the cumulative distribution of the 351 inter-point 
distances $\Delta$  as measured on the plot is shown in figure \ref{kolmos2}, and the average sky separation 
in degrees corresponding to the various separations $\Delta$ on the uniform-exposure plot are shown along 
the bottom of the graph.

 When compared with the inter-point distances found in a large number of sets of 27 points placed randomly on the 
circular plot, the cosmic rays show a large excess of separations at all angles below $30^\circ$, 
clearly non-random, as fewer than 1 in 400 of the random sets match the high number of inter-point distances in 
the range $10^\circ$ -$14^\circ$, related to the size of clumps.  
The cosmic rays are thus clumpy quite apart from what AGNs tell us.  
(An analogous test for randomness was reported in the Auger paper \cite{auger07}, based on inter-event 
distances on the sky, rather than on an equal-exposure plot.)  
To test how well the AGN distribution as a whole accounts for this large-scale pattern, one can construct 
simulated sets of 27 cosmic ray directions by picking AGNs randomly from the catalogue, choosing a ``cosmic 
ray direction'' by applying a Gaussian displacement of rms angle $4^\circ$ (Gaussian displacements 
of $4^\circ/\sqrt{2}$ in two perpendicular directions), and then accepting the direction with a probability given 
by the Auger array detection efficiency at that declination (table \ref{tab-4}).  
There is a complication, as 20\% of the sky is effectively occupied by the galactic zone in which many 
AGNs are hidden, so most of the cosmic rays within $12^\circ$ of the galactic plane are not reproduced in 
such a simulation.  
It seems, in fact, not to change the inter-point distribution greatly, but to check this, a population of 
simulated cosmic rays has been generated by adding ``infilled'' fake AGNs in the galactic exclusion zone, 
by a process of copying random AGNs of rather similar supergalactic latitude and shifting them in supergalactic 
longitude into this zone.  
The distribution of inter-point distances for these simulated cosmic ray directions is shown by the dot-dashed 
line in Figure \ref{kolmos2}, and the effect of reducing the r.m.s. scatter about the AGN directions 
to $2^\circ$ or $0.5^\circ$ is shown by two dotted lines.  
They fit the observed distributions rather well except for a strange deficit, in real cosmic rays, of separations 
near $50^\circ$ -- low gradient of distribution --- with an excess (steeper gradient) near $80^\circ$ (separations 
near 0.7 and 1.0 on the plot). 
There is also one exceptional small separation; and three more around $3^\circ$ may lie outside the average 
distribution, but some of these could occur quite naturally.  
One of the latter is aligned with the very close AGN Cen A (3.4 Mpc), so if Cen A \emph{is} the source, 
deflections could possibly be reduced in this case.  
Also, if these cosmic ray ``deflections'' represent magnetic scattering in a patchy magnetic field, the angle 
between a pair of arriving particles of similar energy from one source can be appreciably less than the typical 
overall deflection, so a simple model of independent deflections will underestimate the frequency of arrival of 
close pairs. 
(Apart from one unusually energetic event, the rms energy spread of the Auger particles is only 15\%.)  
The overall pattern of cosmic rays on the sky thus supports the supposition of their emission from the localities 
of AGNs.  

But this process of selecting randomly the AGNs to serve as source locations ignores the $1/z^2$ geometrical 
attenuation of cosmic ray flux that should occur when deflections are small,
favouring the AGNs closest to us, and also ignores an opposite correction for 
the incompleteness of the catalogue at greater distances.  
Regarding the latter point, in figure \ref{vcvcharac}, the distance 
distribution of the most luminous objects, which would suffer least from a 
flux detection threshold, varies as $dn/dz \propto z$ at low z, as would happen in the case of matter distributed 
uniformly in a thin slab, though the angular spread of AGNs seen in figure \ref{uepp1} indicates that they 
spread much more widely than a slab configuration, so the real $dn/dz$ distribution is probably tending towards a 
more isotropic $z^2$ variation at larger $z$.  
The overall trend of all catalogued AGNs in figure \ref{vcvcharac} is roughly $dn/dz \propto z^{0.3}$, so the 
numbers of catalogued AGNs may have to be multiplied by a factor in the range $z^{0.7}$ to $z^{1.7}$ to correct their apparent space density.
When combined with the cosmic ray geometrical flux attenuation factor $1/z^2$ this suggests 
that the most reasonable distance-weighting factor for selecting ``source AGNs'' would be in the range $z^{-1.3}$ 
to $z^{-0.3}$.  
A distance-weighting factor $1/z$ is a reasonable compromise for the present (stopping the variation 
at $z<0.001$ to avoid infinities), and the use of this factor made a slight improvement to the modelled curve 
in figure \ref{kolmos2}, shown by a full line slightly above the previous ``model'' curve (dot-dash).  
If the catalogue had been supposed equally complete at all distances, and the factor $1/z^2$ thus applied 
(down to $z=0.001$), the very discordant higher curve is obtained, as a few extremely close sources dominate, 
very much at variance with the observations.

The Auger paper showed that where there are cosmic rays of such extreme energies there are AGNs nearby.  
Are there cosmic rays wherever there are AGNs (within 75 Mpc)?  
The overall clumpiness of the cosmic ray distribution has been shown to be consistent with this; 
different source regions are examined next.

\begin{figure}[h!]
\begin{center}
\epsfig{file=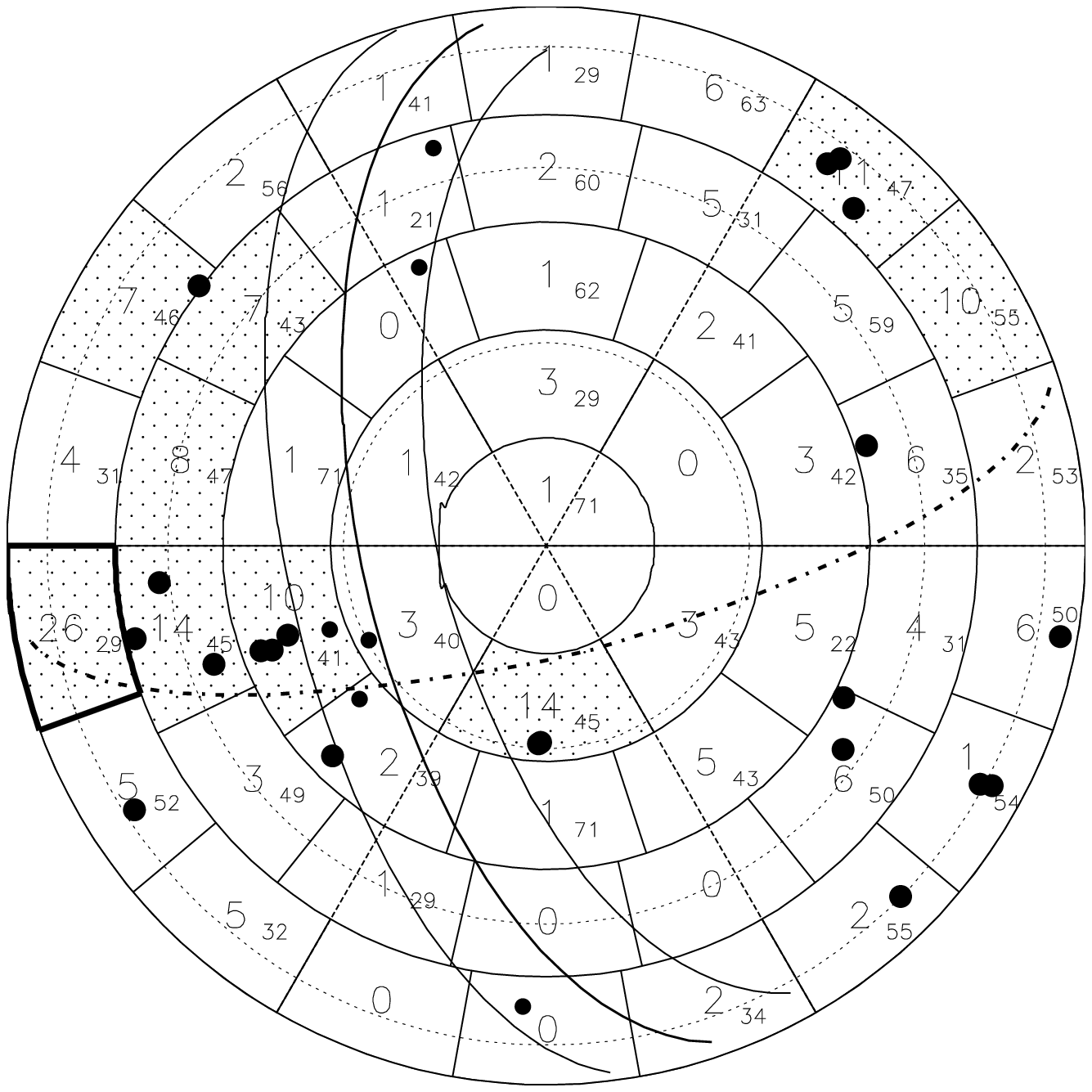,width=0.49\textwidth}
\epsfig{file=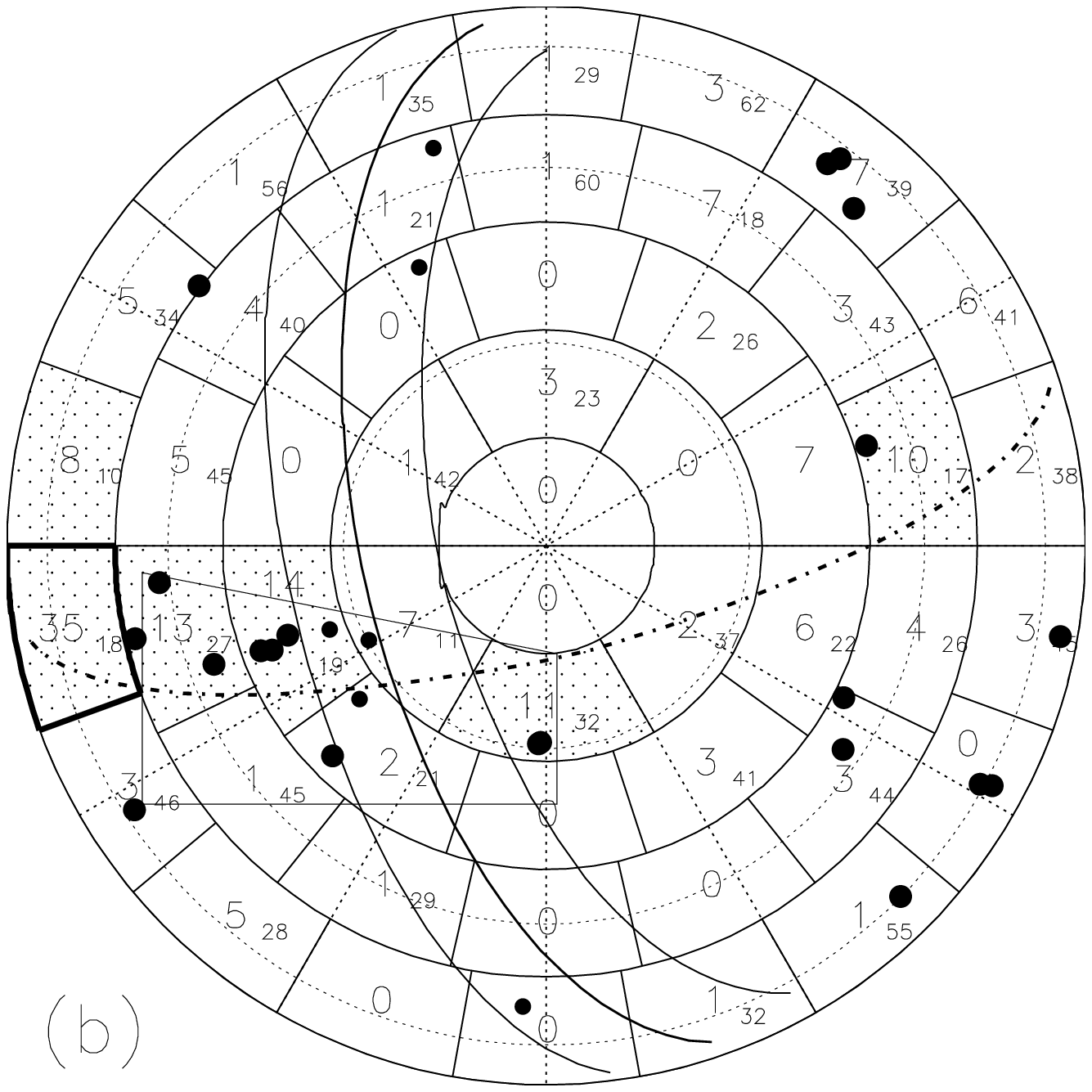,width=0.49\textwidth}
\caption {The uniform-exposure plot divided into 50 segments of equal area.  
(a) The larger numerals give the (rounded) number of VCV AGNs (at a redshift $z<0.018$) 
in each segment, weighted by a relative array exposure factor (made to average 1 for
radii$<$0.75, going to zero at the outer edge).  
Smaller numbers: weighted average distance of these AGNs, in Mpc.  The 9 segments with weighted AGN count $>$6 are shaded. (b) The same, but weight given to each AGN includes an additional factor $\propto 1/z$ (still normalized to 1): see text.  
Here the 6 segments with $>$7 (weighted) AGNs are shaded.} 
\label{ueseg3}
\end{center}
\end{figure}

Figure \ref{ueseg3} divides the equal-exposure plot into 50 segments of equal area, a coarse resolution which 
seems appropriate for this situation.  
(Five annuli, of equal radial thickness, are subdivided into 2, 6, 10, 14 and 18 equal segments in right ascension.)  
To compare the distribution of detected cosmic rays with that of VCV AGNs (with redshifts $<0.018$) as potential 
sources, in figure \ref{ueseg3} each AGN was in version (a) simply given a weight proportional to the detection 
efficiency of cosmic rays from that direction, and the (rounded) total weighted number of VCV AGNs is printed in
each segment, or in plot (b), a distance-dependent factor proportional to $1/z$ was also included in the weight, 
as suggested above.  
In both version (a) and (b), the weights were scaled to give 1 on average within a radius of 0.75 in the diagram.  
In the simple version (a) the weight was then 1.61 in the centre. 

In order to compare cosmic ray and AGN patterns, only the 21 cosmic rays outside the $\pm 12^\circ$ 
galactic exclusion zone will be used: the cosmic rays clearly favour the segments containing many VCV AGNs. 
The numbers in (a) count AGNs when the only weighting is the exposure efficiency (version a), and in the 41 of 
these 50 segments that contain only 0 to 6 (weighted) AGNs, there are 101 of the 208 AGNs and 9 cosmic rays: 
the ratio $CR/AGN$ is 0.09.  
In the whole diagram the ratio $CR/AGN = 21/208 = 0.10$.  
So even where AGNs are sparse, the proportion of cosmic rays is not lower than in the denser regions, 
suggesting in this small available sample that one does not need an exceptionally massive or rare AGN, such as is 
found at the centre of a large cluster, to supply cosmic rays.  
But the heavily-outlined Virgo segment has 26 weighted AGNs (representing a vastly greater actual unweighted number 
in the catalogue) and no cosmic rays, where one might expect about 2.6 ($0.10 \times 26$).
There may be a deficit here, although there is a 7\% probability of recording no counts when 2.6 are expected.  
If the approximate distance-weighting factor (version b) is included, the overall features are not much changed.  
There are now 193 (weighted) AGNs, and 44 segments (those with AGN counts $<$8) contain 102 of the 193 AGNs, 
with $CR/AGN=0.12$ in these, and 0.11 overall.  
But here, the proximity of the Virgo cluster increases its weighting and one expects 3.8 cosmic rays from 
that segment, where none is observed, which does look anomalous.  
In each segment of the plot in figure \ref{ueseg3}, the weighted number of VCV AGNs is given, and also 
(smaller figures) their weighted average distance in Mpc.  
The important segments with more than 6 weighted AGNs and at least one cosmic ray, in case (a) indicate 
45 Mpc as the average distance --- considerably further away than the Virgo cluster (at 16 Mpc) --- whilst the 
smaller $1/z$-weighted averages in case (b) indicate a wide spread in the distance of AGNs in these segments, 
though not necessarily in the distance of the dense active regions.  
Some cosmic rays may originate beyond the 75 Mpc limit of the sub-catalogue used here.  
Two of the 21 cosmic rays have no AGN within the standard Auger window of $3.2^\circ$.  
Is their source further than 75 Mpc, or is the displacement more than $3.2^\circ$?  A technique for extending 
the correlation radius will be proposed in section \ref{rare4sec}.
A more sophisticated treatment of sensitivity of the catalogue at different distances would be appropriate
when more data are available.

The most striking implication of these correlations is that cosmic rays must come from very many sources 
(many $4^\circ$-spread circles being needed to cover the observations) within $\sim 75$ Mpc, and the most 
spectacular strong double-lobe radio galaxies do not supply a large part of them.  
Cen A supplies about $1/4$ of the total extragalactic radio-source flux at 1.4 GHz, with M87 (Virgo A) about 1/5 
of that, and Fornax A about 1/8 of Cen A's radio intensity.  
Biermann et al. \cite{ biermann4} expected these radio galaxies to dominate the cosmic ray flux if their jets 
are powered by the black hole spin, and suggested that intergalactic magnetic fields have displaced the cosmic rays 
from M87 very considerably, to give very many of the observed arrival directions along with the Cen A contribution 
(little displaced because of its proximity).  
A displacement of the Fornax A contribution was also proposed.  
However, if cosmic rays from M87 (Virgo A) were to be redistributed over the region containing the observed 
cosmic rays in the lower left hand quadrant of figure \ref{uepp1}, scattering points randomly over a quadrilateral 
shape containing these observed directions shown in outline in figure \ref{ueseg3}b, only 27\% of the points outside 
the galactic exclusion zone would have a VCV AGN within $3.2^\circ$.  
The observed correlation with AGN positions is too close: such a large-scale deflection is not supported, 
within the present statistics.  
Gorbunov et al \cite{gorbunov} similarly propose a large extension of Cen A cosmic rays, but evidence is given 
in the following sections that magnetic deflections are small.  
Moreover, Gorbunov et al. drew attention to the importance of the huge $1/z^2$ geometrical factor favouring Cen A 
because its distance is only 3.4 Mpc.  
This gives it an advantage of a factor $\sim 200$ compared with the typical VCV AGNs at 40-50 Mpc that appear to be 
correlated with cosmic rays, so Cen A should at least match all the others as a source of local cosmic rays.  
If it does not, we have to suppose that particle acceleration to the highest energies is discontinuous, 
and Cen A is in its off state so far as $10^{20}$ eV is concerned (the jet does not have a non-thermal hot-spot), 
and the same may be true of Fornax A.  
It has been noted that there are two cosmic ray directions within $3.2^\circ$ of Cen A, but these might well come 
from another double-lobed FRI radio galaxy, NGC 5090 (at 48 Mpc) in the list of Nagar and Matulich \cite{nagmat}, 
and not in the VCV catalogue.  
This is within $1^\circ$ of Cen A (see figure \ref{uepp1}).  
The correlation of cosmic rays with these radio galaxies will be examined more quantitatively in 
section \ref{radgal5sec}, but while figure \ref{ueseg3} is before us, it can be seen that there are is a close 
pair of cosmic rays near $\alpha = 341^\circ, \delta =0^\circ$, in a very unremarkable segment of the diagram 
(1 or 0 weighted AGNs).  
The survey by Ghisellini et al. \cite{ghis} also showed few normal galaxies here.  
There are not many AGNs in a further 50\% of redshift either: is there a special object there?  
There is a BL Lac object with extended radio structure, PKS 2201+04, that Nagar and Matulich consider to be 
effectively an extended radio galaxy (see also \cite{moska}), though further away than the others, at 112 Mpc.

But why are no cosmic rays seen from the Virgo cluster region?  
Apart from any contribution from M87, the numerous other AGNs have as yet yielded no visible result, 
as remarked above.  
(a) Is it a technical problem of energy assignment or collection efficiency near the maximum zenith angle of $60^\circ$ ? 
(b) Is the number of AGNs in the Virgo cluster exaggerated by the falling luminosity threshold for nearby AGNs 
in the catalogue, as referred to above? 
(c) Are the cosmic-ray sources not AGNs, but some other objects normally found near AGN but absent in the Virgo cluster ? 
(d) Is cosmic-ray acceleration in Seyfert galaxies suppressed in a closely-packed environment?  
(e) Has the high density of gas generated a larger magnetic field in the supergalactic plane just here 
that deflects particles out of the plane (and out of our line of sight) and hides this source from us?   
Provisionally, it will be noted that option (b) is a distinct possibility, because, as noted above, 
the catalogue appears to have a bias crudely like a $1/distance$ to $1/distance^2$ effect. 
Such a progressive lowering of the power threshold at which AGNs are listed, as one moves to closer distances, 
may eventually move below the effective power threshold needed for cosmic ray generation, so that within some distance 
of Earth, many too-feeble AGNs are included in the catalogue.  
Zaw, Farrar and Greene \cite{farrar} have favoured this explanation, noting that most Virgo cluster galaxies 
have bolometric luminosities below the values typical of other AGNs that correlate with Auger cosmic rays.  
A better view would be obtained from more northerly observatories, which provides an incentive to devise more 
sensitive tests for AGN association.

\section{A right ascension resonance: tracing the decoherence between the patterns of cosmic rays 
and of their sources
  \label{rare4sec}}

Is a $3.2^\circ$ window the most suitable for seeking correlation between cosmic ray arrival directions and AGNs 
at different distances, in different parts of the sky, or at somewhat different energies?  
This is unlikely, whether this dimension arises from magnetic deflections or the size of galaxy or magnetar 
distributions, for example.  
Only the size and particular quality of the Auger data sample permitted a search for the optimal choice in their 
preliminary data set, and the best window size probably arose from its effect in reducing the 
confusion with background objects on the sky.  
For this reason, a correlation test that does not require a particular window size to be selected will 
be proposed, as an alternative technique for exploring the domain where this close association exists.

\begin{figure}[h!]
\begin{center}
\epsfig{file=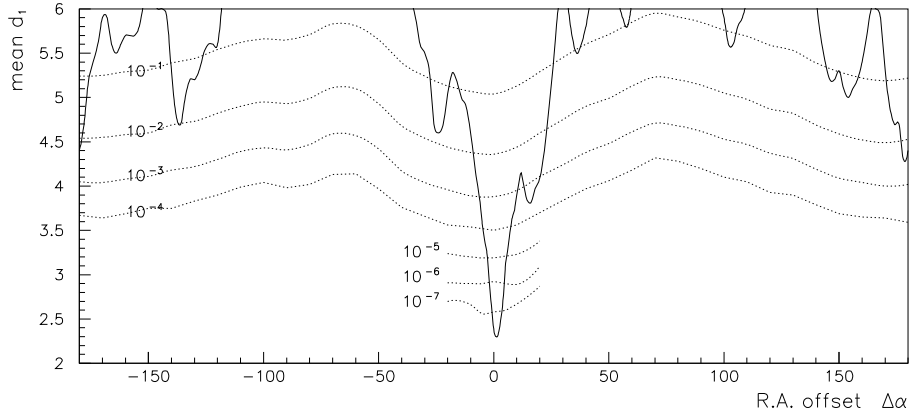,width=12.5cm}
\caption {            
Right ascension resonance plot.  Angular distance $d_1$ of cosmic ray direction from the nearest AGN (z$<$0.018), 
averaged over the Auger set of 21 cosmic ray directions ($|b_{gal} |<12^\circ$), plotted 
when the right ascension, $\alpha$, of each AGN is shifted by $\Delta\alpha$.  
The near-horizontal dotted lines indicate the probability $P(\langle d_1\rangle)$ of finding so small a 
value of $\langle d_1\rangle$ in a set of 21 ``pseudo-random'' cosmic-ray directions 
at a specific offset position.  
(It varies with $\Delta\alpha$ because a $24^\circ$ band almost devoid of AGNs is being shifted across the cosmic-ray pattern: 
only in the unshifted position does it coincide with the belt of omitted cosmic rays.)  
The preponderance of small distances $d_1$ soon disappears when the cosmic ray map 
and the AGN map are displaced slightly.} 
\label{rare4}
\end{center}
\end{figure}

The notable feature of the cosmic ray directions in the original Auger report \cite{auger07} was their 
angular closeness to AGNs.  
Hence a useful measure of association should be the angular distance, $d_1$, between a cosmic-ray direction 
and the AGN nearest to that line of sight in a particular sub-set of the catalogue (such as a particular range 
of redshift).  
Here, $d$ refers to the angular distance in degrees measured on the sky, and not to a distance on the 
uniform-exposure plot.  
Where $\langle d_1 \rangle$ is the average value of $d_1$ for a set of observed cosmic rays, let this be used as a measure 
of their association with AGNs.  
This should be significantly less than $\langle d_1 \rangle_{rand}$ evaluated for a set of the same number of pseudo-random 
directions.  (Here, ``pseudo''-random means that random directions are selected subject to weighting by the 
detector's exposure to each part of the sky.  
They could be chosen using random points on the uniform-exposure plot, for example.)  
But what is ``significant''?  One would formally derive a probability, $P(\langle d_1 \rangle)$,  that a random, or unrelated, 
set of directions would give such a low value of $\langle d_1 \rangle$, by simulating a very large number of ``pseudo-random'' 
sets, but it is hard to be sure that one has not inadvertently constrained the data in such a way as to produce 
a large error in $P(\langle d_1 \rangle)$, so such evaluations are always regarded with some degree of reservation. 
However, if the observed directions were put in the wrong position on the sky by a displacement $\Delta\alpha$ 
in every right-ascension angle, $\alpha$, one would have a set of cosmic ray directions with the same detection 
efficiency, as this is independent of $\alpha$.  
Figure \ref{rare4} shows the average distance $\langle d_1 \rangle$ to the nearest AGN when the relative right ascensions 
of the cosmic rays and the AGNs are shifted by $\Delta\alpha$ ranging from $-180^\circ$ to $180^\circ$.  
As just mentioned, for any value of the closeness measure $\langle d_1 \rangle$ the probability that a set of 21 pseudo-random 
directions will have a value at least as small as this can be calculated by Monte-Carlo simulation, 
and the values of $\langle d_1 \rangle$ at which this probability has the magnitudes $10^{-1}, 10^{-2},.. 10^{-7}$, 
are marked on the diagram, by dotted lines.  
The cosmic rays which are (before shifting $\alpha$) within $12^\circ$ of the galactic plane have not been used, 
as for the vital small shifts $\Delta\alpha$ they cannot be usefully compared with the AGN map.

Only when $\Delta\alpha$ is close to zero is a clear association seen, with a chance probability $< 10^{-7}$ --- a 
sharp ``resonance'' in right ascension. 
Not only is a low $P$ seen, but one knows where it should appear.  
There is much noise, far from the resonance, as a few cosmic rays happen to pass accidentally close to AGNs, 
but the clumpy nature of the AGN pattern results in occasional surprising false correlations, such as is 
seen in the Auger pattern with a displacement near $180^\circ$.
These off-centre troughs are not really significant, as the marked probability $P$ refers to the chance of 
observing such a low value of $\langle d_1 \rangle$ at one particular $\Delta\alpha$, so if a randomly-positioned trough 
is typically about $6^\circ$ wide near its tip, there are about 60 positions available for accidental troughs, 
and multiplying the marked probabilities by $\sim 60$ gives an indication of the chance of seeing so deep a 
trough by accident somewhere: one is very likely to see a purely accidental trough reaching a level of $P =10^{-1.8}$.  
A trough within a very few degrees of the centre is expected, however, so the marked probabilities apply there.

There is a complication, because this rotation in $\alpha$ moves the galactic exclusion belt which contains very 
few AGNs across the pattern of observed cosmic rays.  
It is undesirable to reclassify the cosmic rays as in or out of the galactic exclusion zone, as changes the 
number of directions in the comparison, leading to several problems.  
For this reason, it has seemed preferable to move the AGNs, leaving the cosmic rays fixed, but this still 
leaves the empty belt moving with the shifted AGN distribution (so long as we use the optically-detected 
VCV AGNs), so, as the same set of 21 cosmic ray directions is used at all times, some of the 
``nearest distances'', $d_1$, become appreciably larger for cosmic rays now falling in this belt.  
After testing several alternative remedies, the disturbance that this causes seems to be best treated by limiting 
all individual cosmic-ray $d_1$ values to a maximum or ``capped'' value of $10^\circ$.  
Not only does this suppress the main effect of having a nearly-empty belt, but it also prevents the existence 
of one or two cosmic rays $25^\circ$ from any AGNs (their source perhaps being at higher redshift) from raising 
the overall $\langle d_1 \rangle$ by nearly a degree near the trough of the resonance.  
The movement of the empty belt of AGNs across the sky during this rotation causes changes in the probability 
of observing low values of $\langle d_1 \rangle$ by accident, as shown in the ups and downs of the dotted lines.  
Without the $10^\circ$ cap, this variation is greater.  
In figure \ref{rare4}, $\langle d_1 \rangle$ at resonance is about $2.3^\circ$ for the 21 cosmic rays in unobscured 
directions, whereas the average $\langle d_1 \rangle$ for pseudo-random sets of 21 is $6.8^\circ$, making the associated 
and unassociated populations well distinguished.

To estimate $P$ at a particular shifted position, $\Delta\alpha$, of the AGN pattern,
100,000 ``cosmic ray''points were selected randomly on a uniform-exposure plot, avoiding the 
$24^\circ$ galactic exclusion zone, and the angular distance $d_1$ to the nearest (shifted) AGN
was recorded for each.  Then, a set of 21 samples was drawn from these 100,000, and
$\langle d_1\rangle$ calculated, up to 30 million such sets of 21 being produced,
to find the frequency with which particular low values of $\langle d_1\rangle$ were reached.

\begin{figure}[h!]
\begin{center}
\epsfig{file=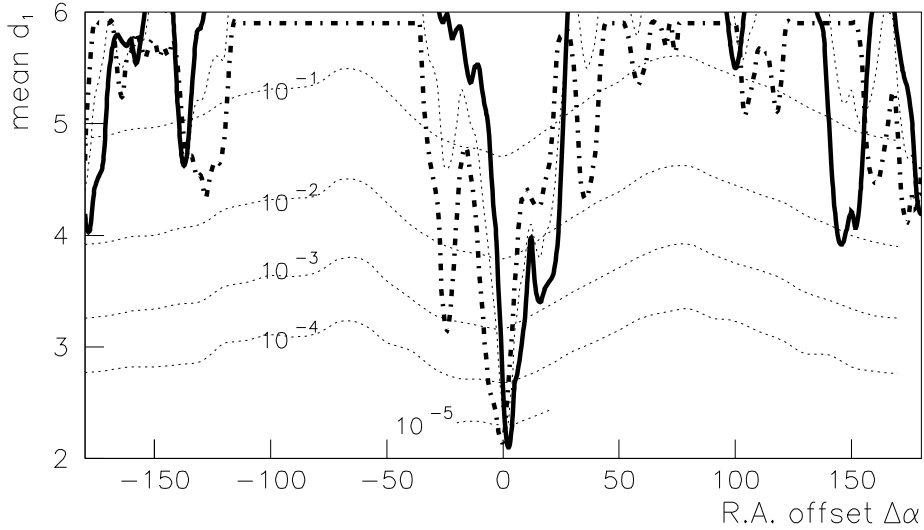,width=12.5cm}
\caption {                       
Right ascension resonance plot for subsets of the Auger data: 
(a) the first 10 cosmic rays $> 12^\circ$ from the galactic plane, used in choosing the selection cuts
(thick dot-dash line), 
(b) the last 11 such cosmic rays, recorded after the cuts had been decided (thick solid line), and 
(c) all 21 such cosmic rays, repeated from figure \ref{rare4} (thin dotted line).  
The dotted lines running across the diagram, showing the probability of reaching so low a value of $\langle d_1 \rangle$ at a 
specified $\Delta\alpha$, apply to the set of 11.  
See figure \ref{rare4} for the probabilities applicable to a set of 21.
The minima reached in subsets (a) and (b) have similar significance levels, in contrast to the 
probabilities of the numbers of $3.2^\circ$ windows containing AGNs which was distorted in set (a) by 
the optimization process.} 
\label{subsets5}
\end{center}
\end{figure}

In the case of figure \ref{rare4}, the marked ``probabilities'' must be regarded with reserve, because the 
selection criteria for the data were decided half-way through the run, in such a way as to maximize the association 
as measured by the number of cosmic rays which had an AGN within a surrounding window of particular (optimized) radius.  
This $\langle d_1 \rangle$ resonance approach removes the relevance of window size, and this should remove the main source of bias, 
but the choice of maximum redshift for the test, and of energy threshold, might still influence the distribution 
in an accidental way in addition to the essential selection of the domain in which a real physical association exists.  
For this reason, figure \ref{subsets5} shows resonance plots for smaller data sets, one of which was taken after 
the selection cuts were fixed.  
The thick dot-dash curve refers to the 10 Auger showers from \cite{auger07} outside the obscured zone of 
low galactic latitude that were observed before the selection cuts were chosen and which were used in making this 
choice; the thick full line applies to the 11 showers detected after this.  
The marked probability levels are for sets of 11 pseudo-random directions, and would be slightly different for 
the smaller first set.  
Even with only 11 observed directions, a $10^{-5}$ probability level is reached (and the same is true for 10). 
In the first data set all 10 of the standard $3.2^\circ$ windows around the cosmic ray directions contained 
VCV AGNs (within 75 Mpc); in the second data set the hit rate was 9/11. 
The nominal probabilities for these hit rates are $10^{-6.1}$ and $10^{-3.5}$, but the first estimate is much too 
small because the selection cuts were chosen precisely to maximize this count, whereas
selection of the cuts does not seem to have 
had such a distorting effect on $\langle d_1 \rangle$.  
The thinner dotted line shows the curve for all 21 events, transferred from figure \ref{rare4}.  
For the nominal probability values $P$ that apply in this case, see figure \ref{rare4}: they are roughly 
the square of the probabilities shown for a set of 10 or 11 directions.  
At $\Delta\alpha$ near $145^\circ$ one can see an accidental dip at around the depth in ``$P$'' ($10^{-1.8}$) 
that was suggested above as likely to occur, but there are some other dips, usually also not significant, 
that are related to the effect of clumps of cosmic rays being moved across a clumpy AGN field, as they recur 
in both sub-sets, and at $\Delta\alpha$ near $-25^\circ$ there is one that is probably partly random.  
The resonance technique could thus detect strong associations with AGN patterns in fairly small data sets, 
even when the best window size is not known.  
The main conclusion from figure \ref{subsets5} is that the association between the cosmic rays reported in the 
Auger paper and the VCV AGNs on a scale of a few degrees are very highly significant, and not generated accidentally 
by the optimization process.

Four other applications of this right-ascension resonance will be considered.  
First: as the technique can reveal such associations in data sets smaller than that of the Auger Observatory,
do associations with AGN directions appear in data of other experiments, viewing the northern hemisphere?   
This raises difficulties which are mentioned below, and the real lessons to be learned are still being explored.  
Second: how significant are the associations with a different list of target objects --- the extended radio galaxies 
put forward by Nagar and Matulich?  
Third: can one detect systematic displacements in the directions that could be caused by magnetic fields?  
Fourth: can one explore how far in redshift the Auger association persists, to check the significance of the 
apparent 75 Mpc limit, and use the position of the GZK horizon to check the energy of the particles?  

\begin{figure}[h!]
\begin{center}
\epsfig{file=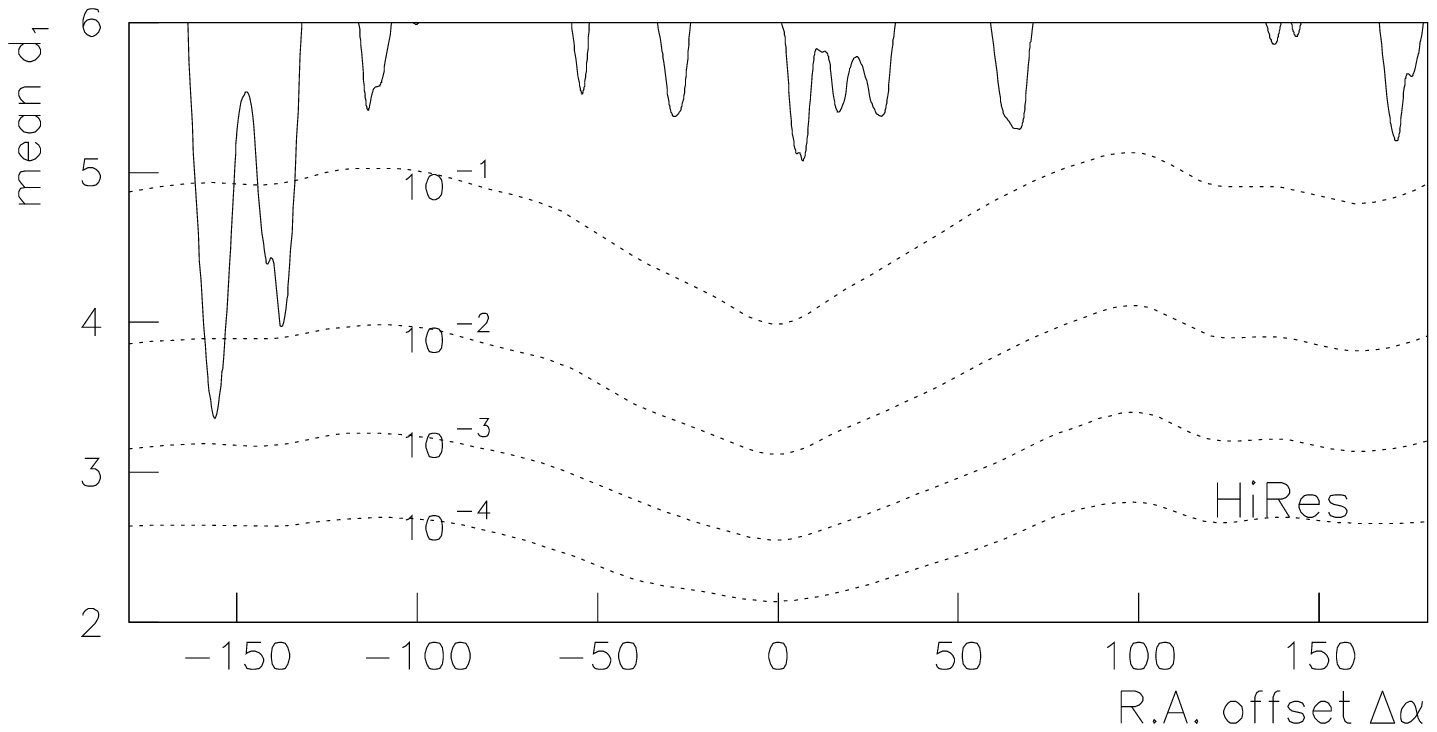,width=12.5cm}
\epsfig{file=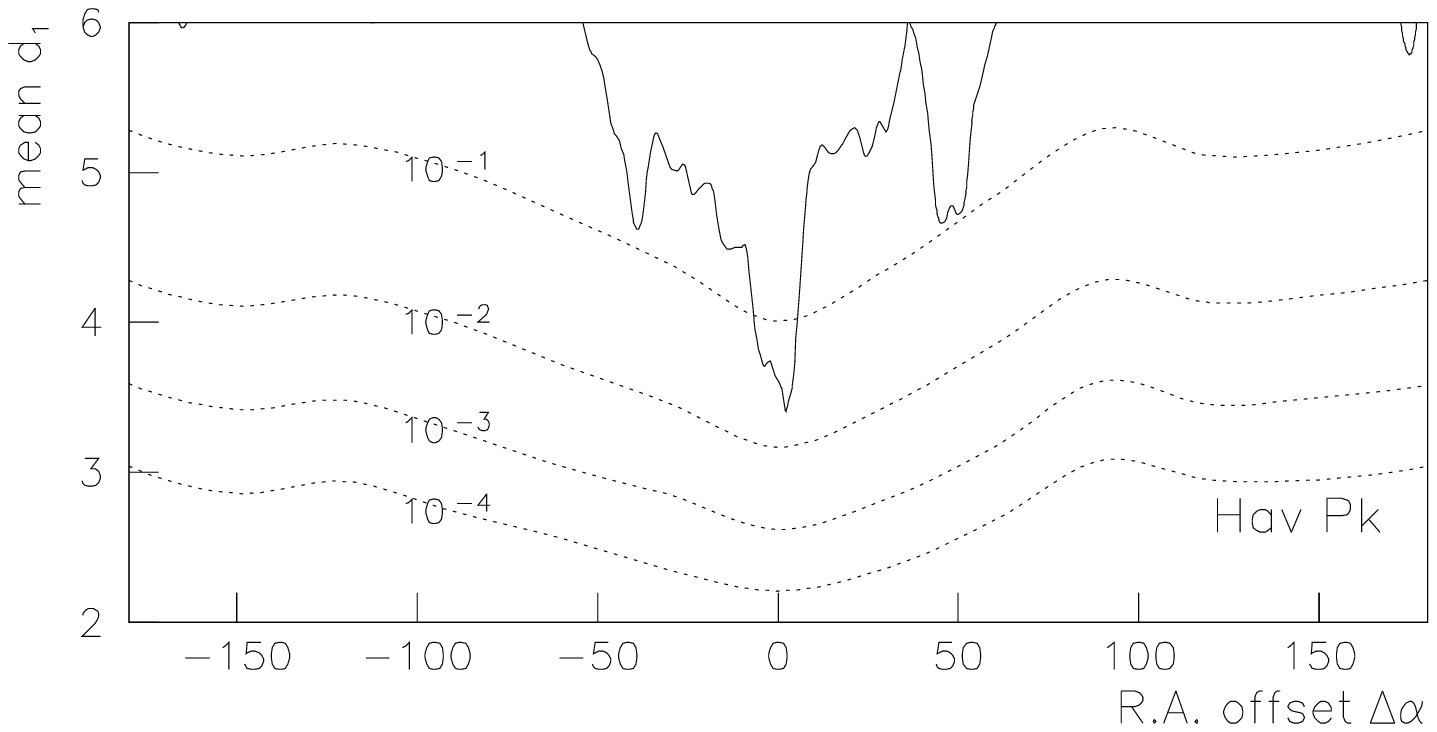,width=12.5cm}
\caption {      
R.A. resonance plots for northern hemisphere observations: cosmic rays $>12^\circ$ from galactic plane.
Upper: HiRes: 11 cosmic rays estimated to have energies above 56 EeV on the Auger energy 
scale \cite{hires,hiresweb}.
Lower: Haverah Park: 11 cosmic rays originally estimated to have energies above 70 EeV,
but possibly about half that.
} 
\label{hiresres6}
\end{center}
\end{figure}

As already remarked, the proportion of Auger cosmic rays to VCV AGNs appears to be lower than usual in the 
region of the Virgo cluster, long assumed to be an important local source.  
The HiRes observatory was well placed to observe this region, but reported seeing no cosmic ray excess in that 
direction, and indeed no association of cosmic rays above 56 EeV with VCV AGNs, using the Auger windows.  
Would some other window size have been more appropriate?   
Abassi et al. \cite{hires} estimated that the energies attributed by Auger to cosmic ray particles were too low 
by 10\% (which is quite possibly an underestimate of the difference, as it depends on the energy at which a 
comparison is made), compared with the HiRes scale, from a comparison of energy spectra, and so they picked out a 
sample of cosmic rays for comparison having energies above ($1.1 \times 56$) EeV on the HiRes scale.   
There were 13 HiRes events supposedly above 56 EeV on the Auger scale \cite{hires,hiresweb}, and a resonance 
plot for the 10 outside the galactic obscuration zone is shown in
the upper panel of figure \ref{hiresres6}.  
The probability estimate is somewhat approximate, being based on the declination distribution of stereo events above 
10 EeV \cite{hiresexpo}.  
There is no signal at $\Delta\alpha = 0$, in conformity with their paper \cite{hires} finding no correlations 
with AGNs.  
The single trough near $-155^\circ$ is a little unusual, though its probability level is near that of the 
peculiar dip seen on the full-sample Auger plots near $180^\circ$, but with a different sample size.  
(The events were displayed on a sky map in \cite{hires}, but the positions used here were those tabulated 
on their web-site \cite{hiresweb}, which seem to be consistent with the map.)  
If a uniform-exposure polar plot is made (as appropriate to the HiRes observing conditions) there is no tendency 
for the cosmic rays to be less numerous in the segments which have few (weighted) AGNs ($z <.018$), unlike what 
was seen in figure \ref{ueseg3}: the cosmic rays here do not follow the AGN pattern
even on a large scale.
Is this astonishing disappearance of any AGN connection a feature of the northern celestial hemisphere --- or 
a surprisingly sharp change with energy that is not fully taken out by the 10\% difference assumed in the energy 
scales used in the two experiments?  
This has to be considered as there must be a rapid change in Auger's sky distribution just below 57 EeV because 
the optimization process found that the significance decreased if the energy threshold was decreased 
despite a very rapid increase ($E^{-3.5}$ integral spectrum) in the number of events available, but a list of 
Auger events below 57 EeV has not been published.

The much earlier experiment at Haverah Park pioneered the particle detector technique used at
the Auger Observatory, and viewed the northern sky. 
Using the energy estimation methods then current, this experiment had 13 showers above
70 EeV, 11 of them outside the galactic exclusion zone (\cite{havpk} with
$\theta_{max}=45^\circ$), and a
resonance plot for these cosmic rays, again matched against the VCV AGNs within 75 Mpc,
is shown in the lower panel of figure \ref{hiresres6}.
There is an apparent resonance with AGNs, with a 2\% chance of being an accident.
But the true energy threshold of these Haverah Park cosmic rays cannot be 70 EeV
because the exposure was only several percent of Auger's:
the energy must be well below the 57 EeV at which the AGN
correlation appears in the Auger observations, and thus very many of these cosmic rays will
originate beyond 75 Mpc, so the dip at $0^\circ$ is less deep.
An interesting feature is that if $3.2^\circ$ windows are drawn around the 11 shower directions, one of them 
contains 12 VCV AGNs (rather than the typical 1.1), being aligned within $1^\circ$ of M87
 at the centre of the Virgo cluster, and if the AGN 
count is extended to 130 Mpc, another window contains 14 AGNs, being aligned 
within $3^\circ$ of a cluster ($\alpha=262^\circ, \delta = 58^\circ$) at 117 Mpc.  
This suggests that the Virgo cluster is not completely inactive as a source, and disfavours the possibility that 
its output is disguised by a large local magnetic deflection.
 
Investigation of the most reliable way to detect associations and estimate energy in such 
datasets is at an early stage, and the result will be reported separately,
but a comparison of these three experimental results suggests that the 57 EeV threshold 
for association with nearby AGNs, found by Auger, may not be simply marking the point at
which interactions with the microwave background radiation limits proton survival 
to about 100 Mpc.  
The depth of shower maximum measured by Auger \cite{augerxmax} indicated that, unexpectedly,
the particles become heavy, and hence highly charged and strongly deflected, as 50 EeV
is approached, though not so below 30 EeV.  
This may be the only way to understand a wide spread of arrival directions in the HiRes
data, if the fluorescence technique favours heavy nuclei rather more than does the array
of water tanks (i.e. assigns them a slightly higher energy).
The Haverah Park array did not find an expected GZK turn-down in the primary spectrum
(as also AGASA, even more strikingly): the apparently most energetic particles were
presumably the badly-measured ones that suffered large fluctuations in their energy assignment: and were probably thus protons.
(The Haverah Park events also occurred in a somewhat lower energy domain, possibly before the highly-charged nuclei had become so important.)
Why should these problems with energy assignment for individual proton showers not
occur also in the Auger data?  
In that case, the domain above 57 EeV apparent energy may be where an accidental tail of
proton showers, which retain good directional information, emerges above the highly
deflected component of heavy nuclei.  The 57 EeV threshold would then be to a large extent
an accident of technique (and an upper limit of the heavy nuclei), and the AGN signal
might be highly dependent on shower selection.  A method for rejecting heavy nuclei on
the basis of the detector signals would be highly desirable.


\section{Correlations with extended radio galaxies
  \label{radgal5sec}}

There is a much smaller alternative catalogue, of large radio galaxies, against which the cosmic rays can be compared, 
which is equally significant, at the one-in-a-million level, may be virtually independent of selection-optimization 
bias, and is undoubtedly plausible, a priori.  
This is the set of 10 ``extended radio galaxies'' listed by Nagar and Matulich \cite{nagmat} as being within the 
declination range observable by Auger and the redshift range selected for the Auger analysis ($z < 0.018$).  
Nearly all are classified as probable FRI radio galaxies, except for Fornax A.  
These authors considered objects included in Liu and Wang's \cite{lizh} catalogue of galaxies with radio jets, 
and selected those for which the radio map showed emission whose total extension on the two sides added up to more 
than 180 kpc (scalar sum).  
Only three of these galaxies are on the VCV list of optically-detected AGNs --- NGC 4261 (3C270), Cen A and Cen B, 
the latter being very near the galactic plane, although there is now no need to exclude the zone of low galactic latitude 
when comparing cosmic rays with radio galaxies, which shine through the dust.  
M87 does not meet their selection test.  
Concentrating first on the statistical argument, of 27 cosmic rays, 4 have at least one of these 10 radio galaxies 
within a ``standard'' $3.2^\circ$ window; yet the average particle from an isotropic flux would have 0.0015 such 
encounter by chance, and there is well below a $10^{-6}$ probability of having 4 ``hits'' in a sample of 27 particles.  
As the selection conditions were not optimized for these targets, this appears to be an essentially independent 
demonstration that the cosmic rays are far from isotropic and truly associated with certain AGN-type objects 
(and the argument would be little affected if one removed the two galaxies that appeared in both lists and had 
nearby cosmic rays).  
The chance of having so many hits by accident is still under 1 in a million if directions are not chosen randomly 
but from a broad region of the sky corresponding to the cosmic-ray pattern in figure \ref{uepp1} but smeared out 
by $30^\circ$ rms spreading (in all cases the acceptance efficiency of the detector is taken into account).  
So the radio galaxies themselves do offer a pattern significance test essentially independent of the VCV association test.
But a $3.2^\circ$ window for counting associations is really too  small for this situation in which the targets 
are less crowded on the sky, as will shortly be seen, so a $\langle d_1 \rangle$ resonance plot is shown, to check this level 
of significance, before moving on.

\begin{figure}[h!]
\begin{center}
\epsfig{file=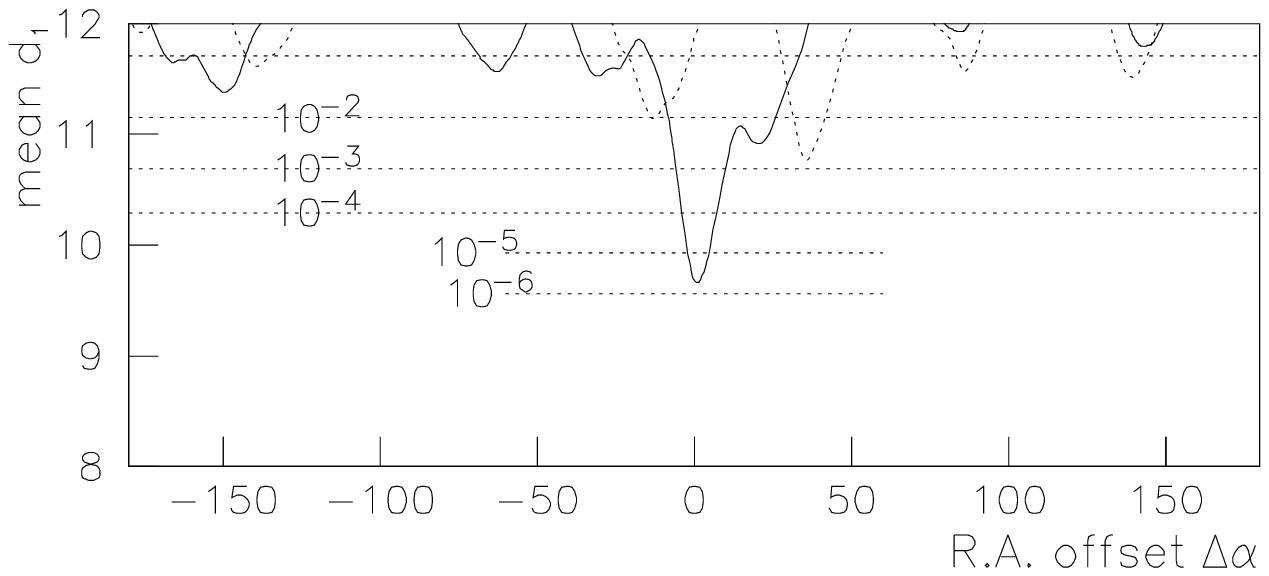,width=12.5cm}
\epsfig{file=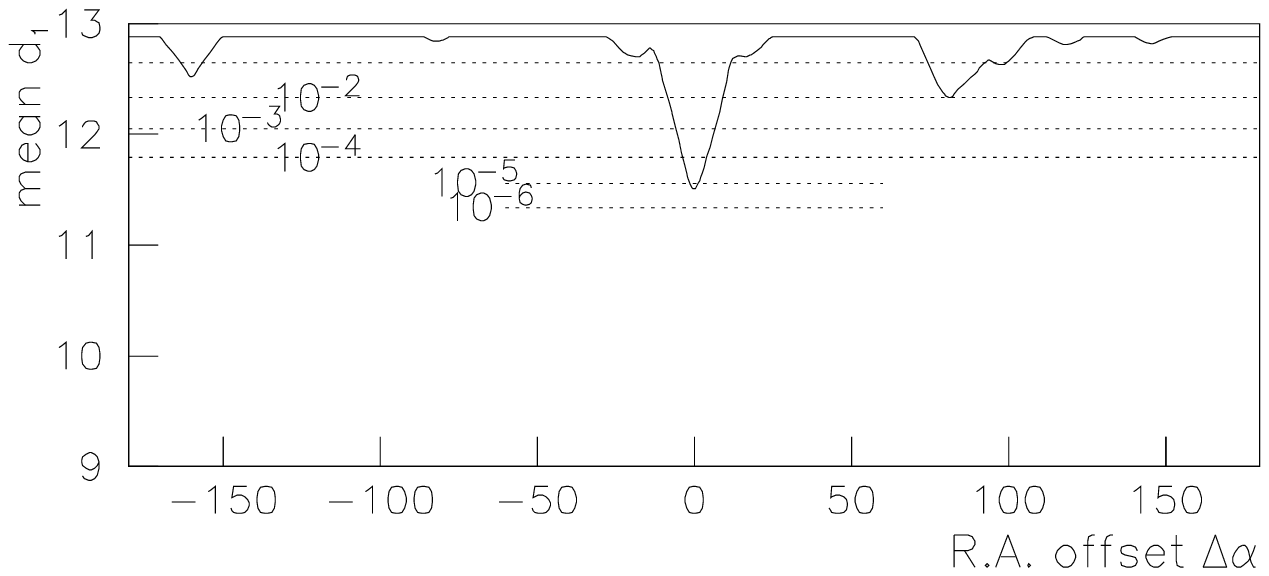,width=12.5cm}
\caption {
Association between the 27 Auger cosmic ray directions and the directions of 
extended radio galaxies listed by Nagar and Matulich.  
The mean distance between the cosmic ray and the the galaxy nearest on the sky is again measured in degrees.  
(a, upper panel) The 10 ``extended'' radio galaxies within 75 Mpc are considered; 
(b, lower panel) the listed radio galaxies (only 2) in the next 50\% range of distance are used.  
The probability lines do not waver in level because there is no special "galaxy avoidance zone" 
moving across the sky now, for radio galaxies; but in case (b) the probabilities should not be relied upon, 
as the unbiased selection of the galaxies in this case is uncertain.  
(a) also shows, dotted, the resonance curve (but not the probabilities) for the radio-jet galaxies 
listed by Liu and Zhang that do not have extension greater than 180 kpc: they do not show a strong signal near $0^\circ$.}
\label{nmres7}
\end{center}
\end{figure}

Figure \ref{nmres7}(a) shows the mean distance in degrees of the 27 cosmic ray directions to the nearest of 
these 10 ``extended radio galaxies'', as the right ascension of the objects is shifted, and 
plot (b) shows the same but in the next 50\% 
extension of the redshift range (0.018 to 0.027),
using the two further radio galaxies that Nagar and Matulich note, 
at positions ($\alpha =52.7^\circ , \delta = -3.1^\circ$ 
on figure \ref{uepp1} , 85 Mpc) and 
at 112 Mpc ($\alpha =331.1^\circ, \delta = +4.7^\circ$).
As these authors do not state whether there were any more 
similar galaxies in this range, there is some uncertainty about the full interpretation of this second plot, 
a subject that will be raised again when considering source distances in section \ref{dist7sec}.  
(Because the density of target galaxies is so low in this catalogue, the random separations are larger, 
and the angular distances $d$ are capped at a value of $13^\circ$ rather than $10^\circ$ in forming the averages 
for these plots, though present results are insensitive to the precise value chosen.)

Without making a choice of window size, the significance of the association (case a) is not quite the same as 
quoted above using a ``standard'' size of window, but this is of no practical consequence. 

These extended radio galaxies were selected from a list of galaxies with radio jets \cite{lizh}.
Are the other jets also effective sources?   Omitting the (few) cases with short jets,
with total length less than 1 kpc, the galaxies selected as having extended lobes comprise about 40\% of the total
in the localities considered here.  A $\langle d_1\rangle$ resonance curve for those that
were not selected by Nagar and Matulich, shown dotted in figure \ref{nmres7}a,
shows no dip at $\Delta\alpha =0^\circ$, indicating that they are not more strongly
associated with cosmic rays than are the typical VCV AGNs.

It may be wondered how one finds room for more sources of these cosmic rays when almost all of them have been 
associated with optical AGNs, above.  
The explanation is, as remarked in section \ref{clustersec}, that the close grouping of AGNs in clusters 
normally prevents identification of what object within a few degrees is the true source, which could be an 
object belonging to the cluster but not in the VCV catalogue.  
Most radio galaxies are such objects, and if they are indeed particularly efficient sources, we may in this case be 
able to identify the actual galaxy that is the source.

The usual $3.2^\circ$ window size must be expected to exclude many cosmic rays if they do typically have rms spreads 
of $\sim 4^\circ$ around the VCV AGNs, and the counting window can safely be extended to $6^\circ$ for greater 
counting efficiency when there is such a small density of targets on the sky: 4 more hits then appear, and 
there is no indication that one is picking up accidental pairings, although there may be a contamination 
from cosmic rays generated by other weaker sources in the surrounding cluster.  
In contrast with the VCV optically detected AGNs, which typically had $\sim 0.1$ cosmic ray associated (see the 
discussion of figure \ref{ueseg3}), the ratio is close to 1.0 now: there are 12 cosmic rays that have at least one 
of these 12 extended radio galaxies within $6^\circ$ (counting out somewhat beyond z=0.018 now), even though about 2 of 
these 12 cosmic rays could be attributed to surrounding VCV AGNs if they are contributing 0.1 each.  
Four of these radio galaxies have no associated cosmic ray (Fornax A, 25 Mpc, NGC 3557, 43 Mpc, NGC 4261/3C270, 
31 Mpc and NGC 4760, 66 Mpc), and it now seems likely that there is a fifth --- Cen A.  
NGC 5090 (at 48 Mpc) has three cosmic rays within $6^\circ$, and so has Cen A (3.4 Mpc) which is so nearly aligned 
that both these galaxies appear in the same events.  
As Cen A has an advantage of $\sim 200$ in its $1/distance^2$ factor relative to most of the other typical emitters 
considered here, but does not overwhelm them, it is probably in an ``off'' state, and the three cosmic rays seemingly 
associated with it are then more logically attributed to NGC 5090.  
Other radio galaxies seem to have two cosmic rays within $6^\circ$ --- Cen B (at 54 Mpc), Markarian 612 (83 Mpc) 
and CGCG 403-019/PKS 2201+04 (112 Mpc) --- though in some cases weaker sources in the cluster may perhaps contribute one.  
I am quoting the distances given by Nagar and Matulich.  
As so many contribute more than one, the 5 undetected suggest that either the emission is not continuous, or that 
the maximum energy is lower for some of them.  
The Nagar and Matulich galaxies should be seen to associate with cosmic rays again in future Auger exposures.

\begin{figure}[h!]
\begin{center}
\epsfig{file=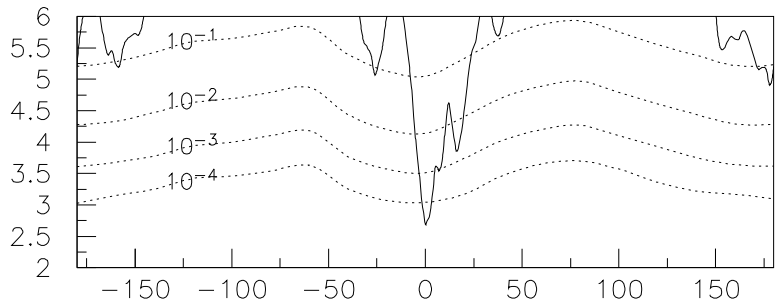,width=12.5cm}
\caption {
Figure \ref{nonnmres8}.  Plot of $\langle d_1 \rangle$ against R.A. shift $\Delta\alpha$ for cosmic rays which do not correlate within $6^\circ$ with 
extended FRI radio galaxies (16 of the original 21 cosmic rays), compared 
with the positions of VCV AGNs after removing H2 galaxies (not true AGNs) and the extended radio galaxies from the list, 
to test a suggestion of residual isotropy by Nagar and Matulich \cite{nagmat}.  
The reduced set of cosmic rays still shows a strong association with the directions of optically detected AGNs.
} 
\label{nonnmres8}
\end{center}
\end{figure}

In \cite{nagmat} it was proposed that these FRI galaxies were the essential sources, and that the remaining cosmic 
rays (perhaps 17 out of 27 from the figures above) were probably not really associated with VCV AGNs, and 
were approximately isotropic.  
The authors also remarked that the correlation with VCV AGNs must be distorted by the inclusion in the catalogue 
of bright H2 galaxies that were originally mistaken for AGNs.  
Their suggestion has been tested by plotting a resonance curve for a reduced cosmic ray sample (having removed 
those which pass within $6^\circ$ of Nagar-Matulich galaxies), and measuring their angular distances from the 
objects remaining in a reduced VCV AGN catalogue for $z < 0.018$ (having removed the 3 specified radio galaxies, 
and also the H2 galaxies as suggested).  
The result, plotted in Figure \ref{nonnmres8}, shows that these cosmic rays that are not seen to be associated 
with the radio galaxies are not isotropic: there is still a pronounced trough close to $0^\circ$, with the chance 
that it arises by accident below $10^{-4}$, still a very significant association.

One merit of the VCV AGN catalogue is that it contains the right density of objects to provide correlations with 
a large proportion of individual cosmic ray directions, but not so many as to make this correspondence unavoidable 
even with random directions, so it makes a striking demonstration of a close association with cosmic rays
numerically possible if it exists.  
It has been argued above that this close association, very surprisingly, does exist, and is very significant even 
when allowance is made for the selection data cuts which optimized this correlation in the first half of the 
data-taking period.  
If one were to look to greater distances, however, source confusion would mount: there would be multiple close 
AGNs for many cosmic rays, confusing the picture quite apart from the need to allow for 
the opposite problem of serious catalogue incompleteness.   
The class of radio galaxy discussed above would be much more suitable than VCV AGNs to trace the extension of 
cosmic ray sources beyond 75 Mpc, because their number density is low enough to avoid source confusion, 
and the catalogue will not suffer from incompleteness until greater distances are reached.  
The X-ray selected AGNs in the Swift BAT catalogue can also be useful in this way, and George et al. \cite{swift} 
have already suggested that a correlation appears to extend to about 100 Mpc.

Thus it appears that both FRI radio galaxies and Seyfert galaxies can accelerate protons to $10^{20}$ eV, though the 
latter have a lower output.  
It seems that an intense hot-spot, as seen at the termination of FRII jets, and anticipated as a promising 
acceleration site \cite{araa,biermann4}, is not implicated.  
If these two classes of active galaxy make up the relevant accelerators, the acceleration site may possibly be 
nearer to the base of the jet.

\section{Magnetic deflections
  \label{mag6sec}}

\begin{figure}[h!]
\begin{center}
\epsfig{file=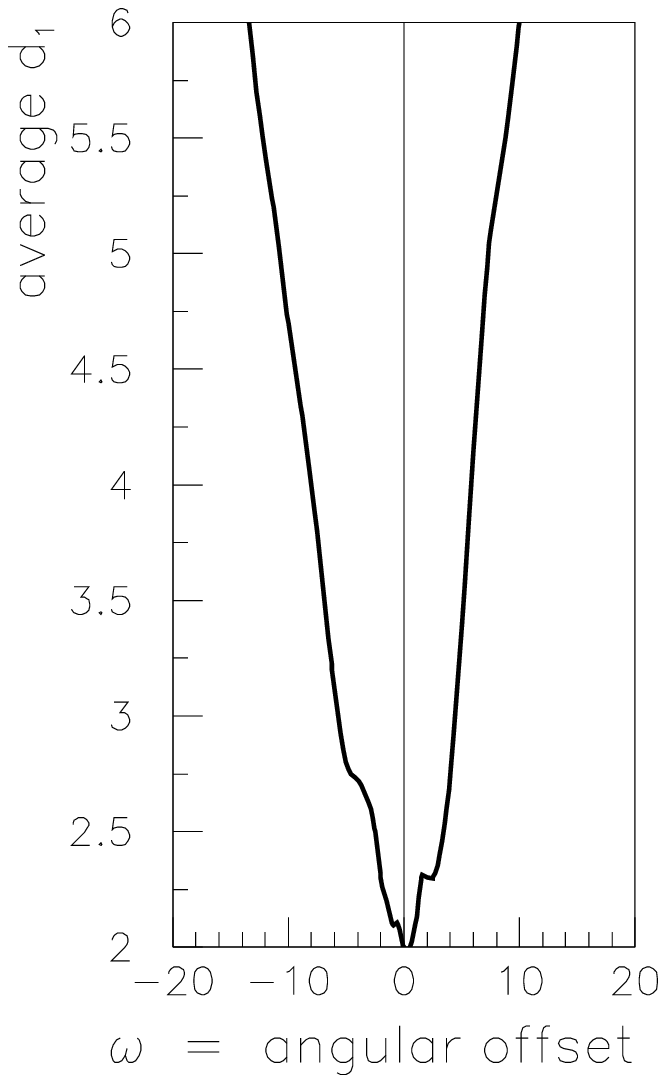,width=0.32\textwidth}
\epsfig{file=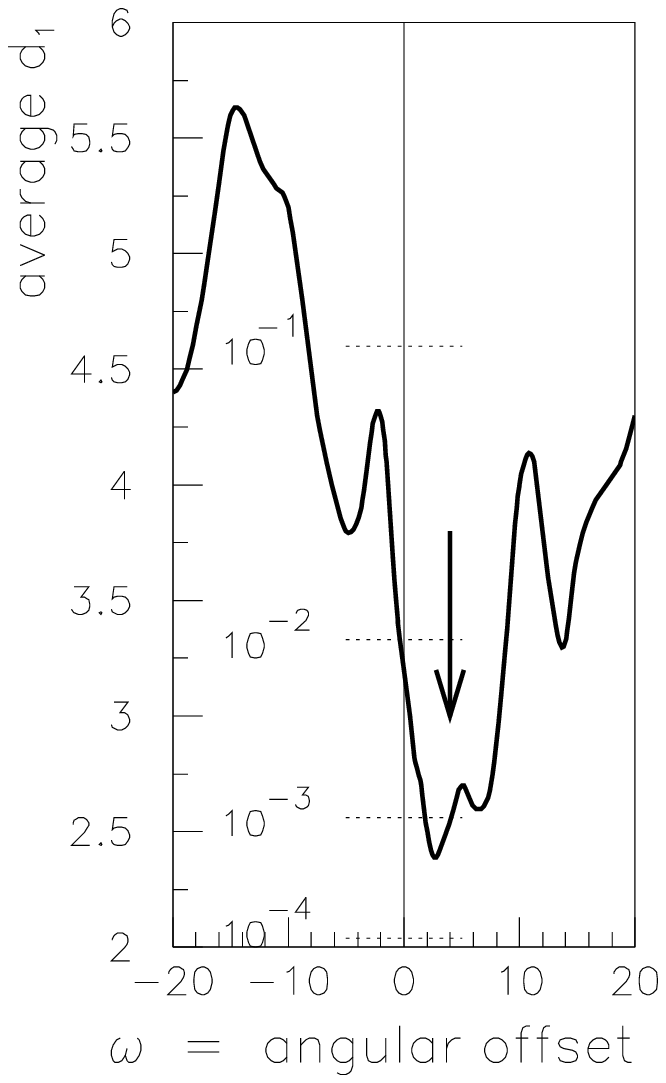,width=0.32\textwidth}
\epsfig{file=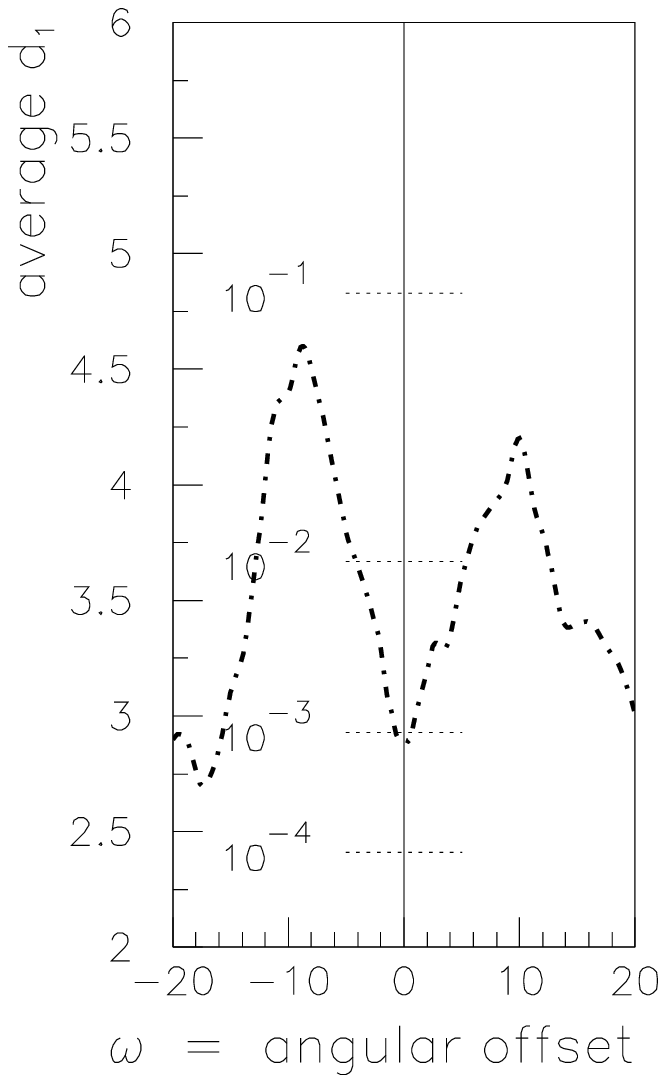,width=0.32\textwidth}
\caption {Position of the resonance between Auger cosmic-ray directions and the AGN positions ($z < 0.018$) when 
the AGN patterns are shifted in a direction related to galactic coordinates: 
in (a) and (b), the AGN positions are shifted by an angle $\omega$ parallel to the galactic plane, and in 
(c) by an angle $\omega$ perpendicular to the plane (see text).  
(a) uses the 12 cosmic-ray directions with galactic latitude, $b < -12^\circ$; 
(c) uses the 9 which have $b > 12^\circ$. 
(b) uses 7 of the latter 9, which arrive from a relatively limited part of the sky around the direction 
($l = 308^\circ, b = 24^\circ$).  
In case (b), where the particles have passed just above the galaxy, they appear to have been deflected by 
about $4^\circ$, as though by a $B_z$ component of the galactic magnetic field.} 
\label{magres9}
\end{center}
\end{figure}

If $3^\circ$ of the spread of proton arrival directions around their sources represents a deflection in varying 
magnetic fields in travelling $\sim 45$ Mpc, an effective transverse magnetic field $\approx 0.1 \mathrm{nG}$ would produce this.  
If the field changed randomly on a scale of $L$ Mpc, a typical total field strength in each domain of  
$\approx 1.6 \mathrm{nG} /\sqrt{L}$  would be required.  
This must be an upper limit, as there will be other factors contributing to the scatter.  
A systematic deflection caused by a net average field normal to the line of sight might be detected by a shift in 
the position of the ``right ascension resonance'' from its expected position at $\Delta\alpha =0$ --- if the field 
had a component parallel to the earth's rotation axis.  
But other field directions can be investigated by shifting the AGN pattern in some other direction on the sky by 
choosing a different axis for the pattern rotation.  
The resonance effect has been examined when the ``AGN sky'' is rotated through small angles about the galactic pole 
or about an axis perpendicular to this (i.e. in the galactic plane, chosen to be at a galactic longitude of 
$45^\circ$ so as to be nearly perpendicular to the main concentration of cosmic-ray directions).  
Of special interest is the set of AGNs at positive galactic latitude, seen on the left hand side of figure \ref{uepp1}, 
for these 9 at galactic latitude $>12^\circ$ are mostly grouped in a small part of the sky and so may travel through 
a similar magnetic field.  
On rotating the AGN sky about the special axis normal to the galactic pole, which is also about $90^\circ$ from the 
majority of these 9 directions, no perceptible shift from zero in the position of the resonance was seen 
(Figure \ref{magres9}c), although this plot is very noisy, 
and a much larger data set may be needed to make clear what is happening.  
But if a rotation about the galactic pole is made, thus shifting the AGNs parallel to the galactic plane, 
they are seen to need an offset in order to match the cosmic-ray directions.  
To make the sampled region of sky more well defined, two outlying cosmic rays may be omitted (one at much smaller 
galactic longitude than the others, and one at the high latitude of $54^\circ$), leaving a more compact 
group of 7 arrival directions whose centroid is at galactic longitude $308^\circ$ and latitude $24^\circ$.
The resonance curve (figure \ref{magres9}b) shows that the AGNs best match the cosmic ray directions when offset 
by about $4^\circ$ parallel to the galactic plane. 
(A shear was applied in these rotations, as described in the next paragraph.)  
This sky rotation about the galactic pole tests any deflection of cosmic rays parallel to the galactic plane 
(due to a ``$B_z$'' field component, perpendicular to the galactic disc), whilst the previously mentioned rotation 
tests any displacement perpendicular to the galactic plane, due to a field component parallel to the plane 
(though cylindrical symmetry of the galaxy may tend to make the averaged field along the line of sight nearly 
cancel out in this case, where no displacement effect was found).  
No clear offset was seen if the other 12 (negative latitude) cosmic rays were used (figure \ref{magres9}a).
These directions are more widely scattered, and typically further from the galactic plane, where the average 
galactic field strength might be considerably smaller. 

A rotation of the AGN pattern about any axis, such as the galactic polar axis, produces a smaller actual displacement 
on the sky as the pole is approached.  
Since the actual displacement of cosmic rays was of prime interest in this part of the investigation, 
a ``shear'' was applied to the rotations described in this section: 
any AGN was displaced in longitude, $\lambda$, by $\Delta\lambda  = \omega .\sec\phi$, where $\phi$ is the latitude, 
rather than all by the same $\Delta\lambda = \omega$, thus having the same actual shift $\omega$ on 
the sky for each AGN, making it easy to read off the displacement of the particle. 
The terms ``longitude'' and ``latitude'' here relate to the particular axis of rotation that is being used.  
(A rather similar curve, but without such a large sideband noise is obtained in the case \ref{magres9}(b) by 
rotating in right ascension instead of in galactic longitude, as in this region of the sky the two directions 
are not very different.)

Such a deflection of $4^\circ$ by the galactic magnetic field would be quite compatible with the range of 
expected values shown in figure 8b of Abraham et al.\cite{auger07}, and would correspond to a field directed 
``down'' towards the galactic disc, with a strength integrated along the particle's path of 
$6 \mu\mathrm{G} \times\mathrm{kpc}$, if the particles are protons --- say, 
$0.3 \mu\mathrm{G}$ on average over 20 kpc.  
The trajectory passes about 2 kpc above the galactic plane at about 7 kpc from the Galactic Centre.  
Since deflections, at a known energy, will be proportional to nuclear charge, the particles are surely not 
very much more highly charged than helium nuclei, and at this energy are very unlikely to be helium nuclei, 
as these survive only about 1 Mpc in the microwave background.  
Of course, not having enough data to treat several different parts of the sky separately, the magnetic field 
cannot be separated from a metagalactic field of $6 \mathrm{nG} \times\mathrm{Mpc}$.  
There may well be a significant metagalactic contribution seen in the widespread random $3-4^\circ$
scatter discussed previously, as the systematic shift was only seen for a particular 
set of directions, at the rather low galactic latitude of $24^\circ$.  
The value of extending these tests to a larger data set is clear.

A $4^\circ$ displacement of the resonance of course implies some changes in the list of which AGNs appear in 
$3.2^\circ$ windows around the cosmic ray directions.  
With this special group of 7 cosmic ray directions there are still 2 empty windows (to 75 Mpc),
but the number of AGNs in the 5 active windows is increased --- so it may not yet be
convincing that there must be a magnetic deflection.
There now appear 9 AGNs in the standard windows around revised ``undeflected'' cosmic ray directions, 
instead of 5 in the original list.  
Cen A now appears in only one window rather than 2, but one of the new AGNs brought in is the FRI AGN IC 4296 on 
Nagar and Matulich's list \cite{nagmat}, which now has a distance $d$ of $0.4^\circ$ instead of $3.8^\circ$.

\section{Correlations with AGNs in different distance ranges
  \label{dist7sec}}

The Auger AGN study \cite{auger07} found that the correlation was most significant when only the AGNs having 
redshift $< z_1 = 0.018$ (distance $<75$ Mpc) were used.  
This would have important implications, discussed in the next section.  
However, a maximum significance with a redshift limit of $z_1$ does not mean that there are few cosmic rays 
from beyond this limit; so direct information on the distances of travel of the detected cosmic rays should be sought.  
Three approaches to this tricky problem will be sketched here.

(1)  Looking for coincidences (within $3.2^\circ$) with VCV AGNs at different distances is not helpful in a very 
straightforward way, as these objects are so numerous that a typical cosmic ray direction eventually matches AGNs 
at different distances.  
This is shown in table \ref{tab-3}, where, in the line labelled ``z .000-.017'', the 21 cosmic rays outside the galactic obscuration 
zone are shown in the sequence of their detection, with ``x'' showing that the particle matched an AGN in this distance 
range (i.e. the cosmic rays discussed previously), and ``.'' signifying that there was no match: there were 19 ``hits''.  
The following two lines similarly record whether these same 21 cosmic rays matched AGNs in the next succeeding 
ranges of redshift, each 50\% as large as the first.\\

\begin{table}[!h]
\begin{center}
\begin{tabular}{l}
\hline
\texttt{AGN redshift}\\
\ \ \texttt{.000-.017:   x x x x x x x x x x | x x x .\ x x x x x x .\ \ \ \ 19}\\
\ \ \texttt{.018-.027:   .\ .\ .\ .\ .\ .\ .\ x x .\ | .\ x .\ x .\ .\ .\ .\ .\ x .\ \ \ \ \ 5}\\
\ \ \texttt{.028-.035:   .\ .\ .\ x x .\ x .\ x .\ | x x .\ .\ x .\ .\ x x .\ x\ \ \ \ 10}\\
\hline
\end{tabular}
\end{center}
\caption{Sequence of cosmic rays matching AGNs (see text).}
\label{tab-3}
\end{table}

The point of origin of the particle is obviously not determined unambiguously: a line of sight can pass within 
$3.2^\circ$ of more than one AGN.  
The selection ``cuts'' were chosen so as to optimize the number of hits in the first half of the run --- before the 
dividing line in the sequences shown above --- and the 10/10 hit rate in the initial period is hence distorted.  
The 9/11 hit rate after this is close to what would be expected if cosmic rays do arrive from directions having 
$3.5^\circ$ rms scatter around VCV AGNs: it was found that 77\% of particles should then have an 
AGN within $3.2^\circ$ along the line of sight.  
Such simulations also show that 21 particles originating in the near zone, $z<0.018$, would typically score 12 hits 
in the second and third distance zones, not very different from the 15 recorded, so that there are probably not 
very many particles originating in these more distant zones (though several
associations could be missed because the VCV catalogue will become significantly incomplete there).  
A more elaborate study of this limited sample suggests that
(because of the high hit-rate inthe first line) most particles originated within 75 Mpc, but very crudely 
0-25\% might come from sources beyond 75 Mpc, though a more precise estimate cannot be made.

\begin{figure}[h!]
\begin{center}
\epsfig{file=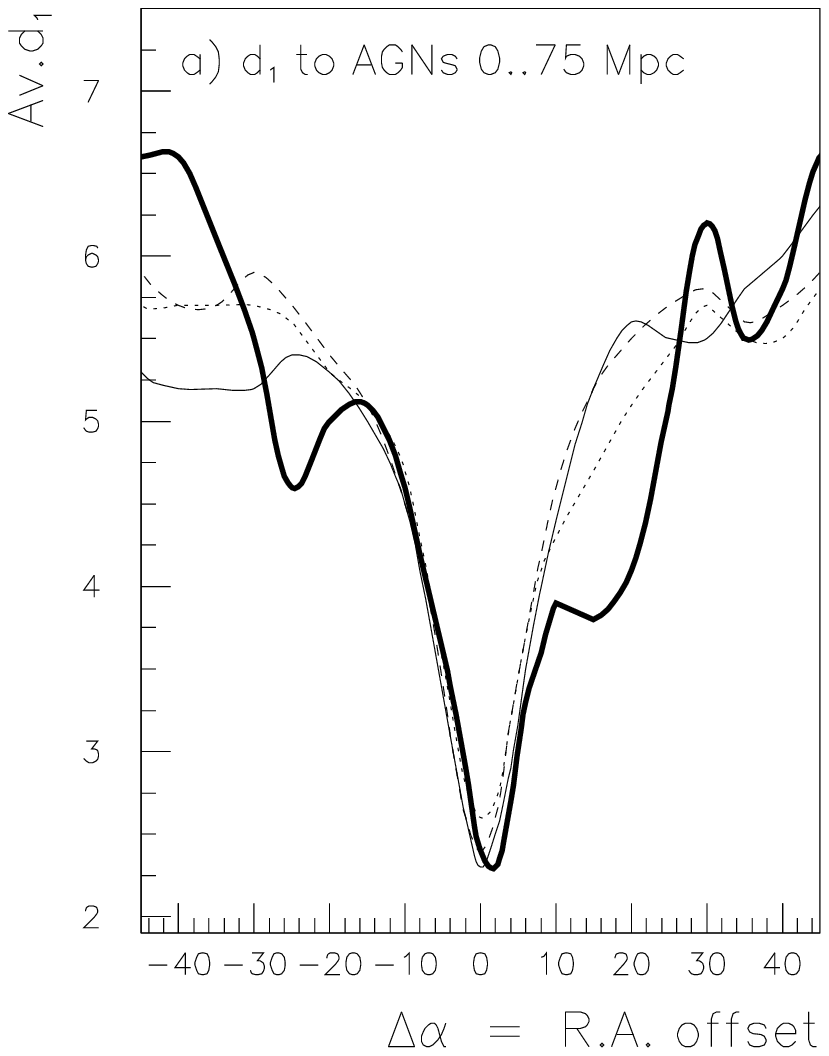,width=0.32\textwidth}
\epsfig{file=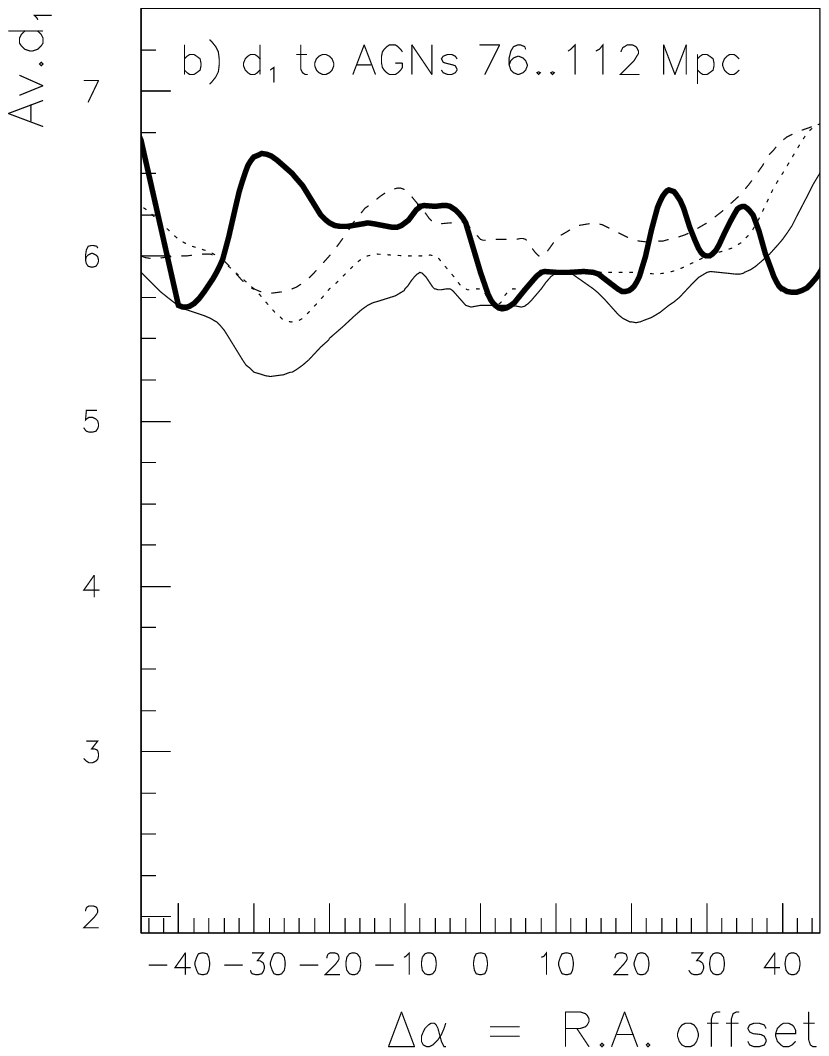,width=0.32\textwidth}
\epsfig{file=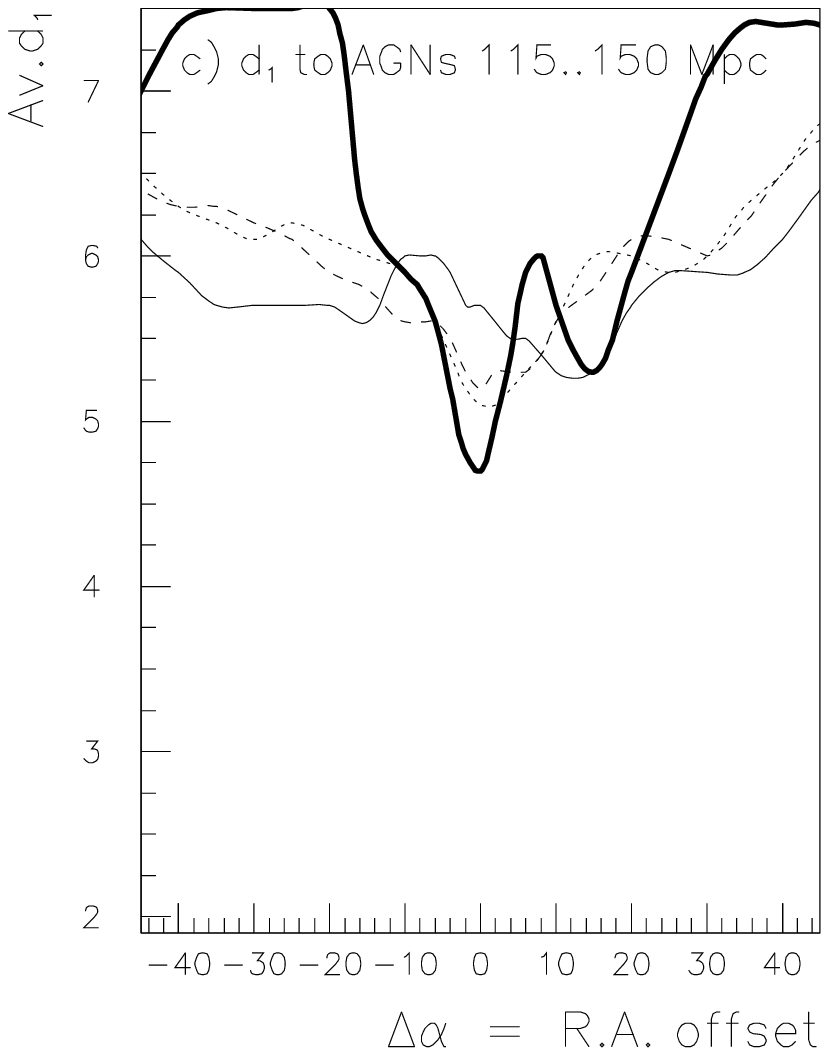,width=0.32\textwidth}
\caption {Right-ascension resonance plots for the 21 Auger cosmic ray directions not near the galactic plane.  
The mean distance to the closest VCV AGN $\langle d_1 \rangle$ is plotted against shift $\Delta \alpha$ in right ascension of 
AGNs by the thick lines, (a) for AGNs in redshift range $<0.018$, (b) 0.018-0.027, (c) 0.028-0.035. 
The thinner full, dashed, dotted lines show the average curves found in models in which the cosmic rays are 
emitted from the vicinity of VCV AGNs, with detectable output reduced by a factor falling to zero at redshift 
$z_{max}$ in the range 0.018 to 0.025 (see text). 
When $z_{max}$ exceeds 0.024, approximately, the correlation with the closer AGNs, noted by the Auger report, weakens. 
The seeming correlation with AGNs at $z>0.027$ appears to stem in part from accidental alignment of AGN clusters.}
\label{3zores10}
\end{center}
\end{figure}

(2) Resonance plots can also be used to study the extent of the correlation with increasing distance, as they are not dependent on window size, though the plots are disappointing.  
Right ascension resonance curves for the 21 Auger cosmic rays are shown in figure \ref{3zores10} 
for $\langle d_1 \rangle$ using angular distances to the nearest AGN listed in particular redshift ranges of the VCV catalogue: 
(a)  $z<0.018$, (b)  0.018-0.027, (c) 0.028-0.036.  
It is at first disconcerting to see that although most of the published cosmic ray directions (in unobscured 
regions) align closely with AGNs within 75 Mpc, as discussed above, and as shown in figure \ref{3zores10}a, 
they also have a (much less significant) alignment with AGNs between 116 and 150 Mpc (figure \ref{3zores10}c).  
Although this is somewhat like the results for window hits listed above, it is surprising that the 116-150 Mpc group 
shows a small apparent resonance near $\Delta \alpha =0$ (that was not seen in the simulated curve, below).  
To extract the information on distance of origin, the best that can be done at present is to model the correlations 
one would get if the observed cosmic rays originate in a specific range of distance.  
So, for this purpose, I have assumed that cosmic rays originate as though from any VCV AGNs, are scattered 
with a characteristic spread, and that at greater distances a decreasing fraction of the AGN's output is seen, 
because GZK energy losses allow us to receive only those particles emitted above an energy threshold which is 
increasingly above 57 EeV for more distant sources.  
The Auger sample has an energy spectrum above 57 EeV which may be represented by  $j = (E_{max} - E)/(E_{max} - E_o)$, 
where $j$ is the fraction of particles above energy $E$, $E_o$ = 57 EeV, and $E_{max} \approx 92 EeV$.  
This ignores a single particle of energy well above 92 EeV, which is of no consequence in this context.

Assuming the local spectrum does not vary much with position in space, it might be treated roughly as though
$E_{max}$, above, is the maximum energy with which most particles are generated, and that particles cannot be received 
here from sources more distant than $D_{max}$ (redshift  $z_{max}$), corresponding to the distance over which 
protons lose an amount $\Delta log_{10}E = 0.22$ (falling from 92 to 57 EeV).  
This effect is modelled approximately by multiplying the output of any AGN at redshift $z$ by a factor 
$(1-(z/z_{max})^2)$ to represent the decreasing fraction of its particle output that remains above 57 EeV by the time 
it reaches us.  
(``92 EeV'' was quoted merely in explanation: no energy value is used at this stage, but only a parameter $z_{max}$.)

Varying  $z_{max}$ (and the rms spread of cosmic rays around source AGNs) to get a crude fit to the resonance trough, 
$\langle d_1 \rangle$ near $\Delta \alpha = 0$ in the three redshift zones out to 0.036, the best fit was with 
$z_{max} \approx  0.023$ and slightly favours a scatter varying with $\sqrt{z}$ (rms being $3.5^\circ$ at 45 Mpc) rather than being constant.  
Models with $z_{max}$ = 0.018, 0.0216 and 0.0252 gave the resonance curves shown by various thin lines in 
figure \ref{3zores10}.  
The present rough models and low statistics cannot give accurate fits, but if $z_{max}$ is increased further the 
trough becomes less deep in the comparison with AGNs in the nearest redshift zone, $<0.018$ (figure \ref{3zores10}(a)), 
that was picked out by the Auger authors.  
A range zmax of 0.023 corresponds to a maximum source distance of 96 Mpc if $H_o = 72$, a travel time of close to 300 Myr. 
When better statistics are available and refitting is attempted, the distance-dependence of the selection bias for 
VCV AGNs should be investigated, although it would be unlikely to have much effect on a ``cut-off'' point.

\begin{figure}[h!]
\begin{center}
\epsfig{file=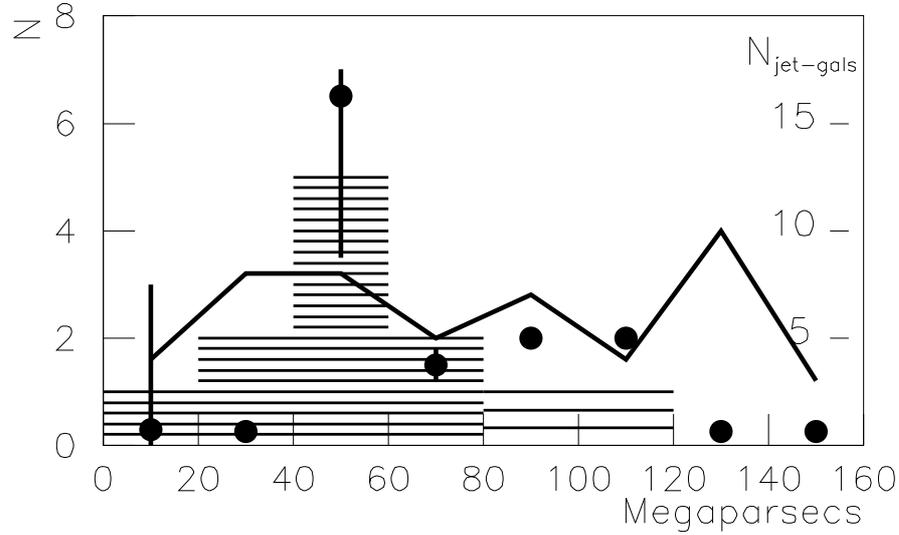,width=12.5cm}
\caption {The histogram shows the numbers of extended radio galaxies as defined by Nagar and Matulich in 20 Mpc intervals.  
Beyond 75 Mpc the galaxies may have been chosen more selectively (see text).  
The line shows the number of galaxies with radio jets of at least 1 kpc \cite{lizh} 
--- in the field of view of the Auger Observatory, but without selection for lobe extension --- the right hand scale 
applies to these.  
The black points are the numbers of cosmic rays that appear to be associated with these radio galaxies, 
none in fact being associated with non-extended galaxies.  
3 cases attributed to NGC 5090 could alternatively be attributed to Cen A: the vertical lines show the changed 
counts that would result. }
\label{erghist11}
\end{center}
\end{figure}

(3)   The alignment of cosmic rays with extended radio galaxies offers a much better chance to investigate the source 
distances, as multiple alignments for a cosmic ray will rarely occur, apart from the special case of Cen A and NGC 5090.  
The shaded histogram in figure \ref{erghist11} shows the numbers of such galaxies (as listed by Nagar and Matulich) 
found in 20 Mpc intervals of distance, but the lightly shaded part reaching as far as 120 Mpc contains objects
that seem to have been added 
because they were seen to align with cosmic rays, and may not be a complete sample in this region.  
To indicate how the radio-galaxy distribution may be expected to continue, the total number of galaxies with radio jets, listed by 
Liu and Zhang \cite{lizh}, excluding those with total jet length less than 1 kpc but without any selection based 
on extension of lobes seen in radio images, is shown by the line, the scale being shown on the right.  
Below 80 Mpc, the extended objects amount to about 40\% of this total, and the scale has been chosen so that the 
line may thus indicate how the numbers of extended radio galaxies may be expected to continue beyond this distance.  
The filled circles show the numbers of cosmic rays apparently associated with the radio galaxies in each 20 Mpc 
--- within $6^\circ$ as in section \ref{radgal5sec}.  
(The ``error bars'' show the effect of changing the source attribution from NGC 5090 to Cen A in 3 cases.)  
All these radio galaxies with jets have been included in these counts of associations, but only the extended 
(mostly FRI) radio galaxies mentioned by Nagar and Matulich showed an association.  
Thus the cosmic rays may indeed all originate within about 120 Mpc, though some sources may have been missed if 
the catalogue of Liu and Zhang is incomplete in the range 120-160 Mpc.  
The source peak near 50 Mpc is notable.

Hence, this treatment roughly confirms the $z_{max}$ figure from the Auger optimization: it is probably below 120 Mpc 
for the set of particles that still have energies above 57 EeV when they arrive.  
This must be controlled by their maximum energy on production, and the rate at which they lose energy.

\section{Implications of very local sources: a cut-off in the source spectrum? 
  \label{implic8sec}}

If one receives few or no particles above $E_{thresh}$ which originated beyond 100 Mpc, the energy of the particles 
released from the source must not be too high, and the rate of loss of energy, primarily through interactions with the 
cosmic microwave background radiation (CMBR), must be rapid, so that the energy falls below $E_{thresh}$ within 100 Mpc.  
The rate of energy loss in the CMBR varies rapidly with proton energy, and it seems that the loss would be sufficient 
to drop below ``57 EeV'' within this distance only if, firstly, the particle energies are higher than quoted in the Auger paper, 
presumably because of remaining uncertainties in the fluorescence efficiency of air; 
and, secondly, the fall of the energy 
spectrum of cosmic rays must be steeper than that imposed by these energy losses: the observations must be closely 
approaching the upper energy limit of the accelerators.  

\begin{figure}[h!]
\begin{center}
\epsfig{file=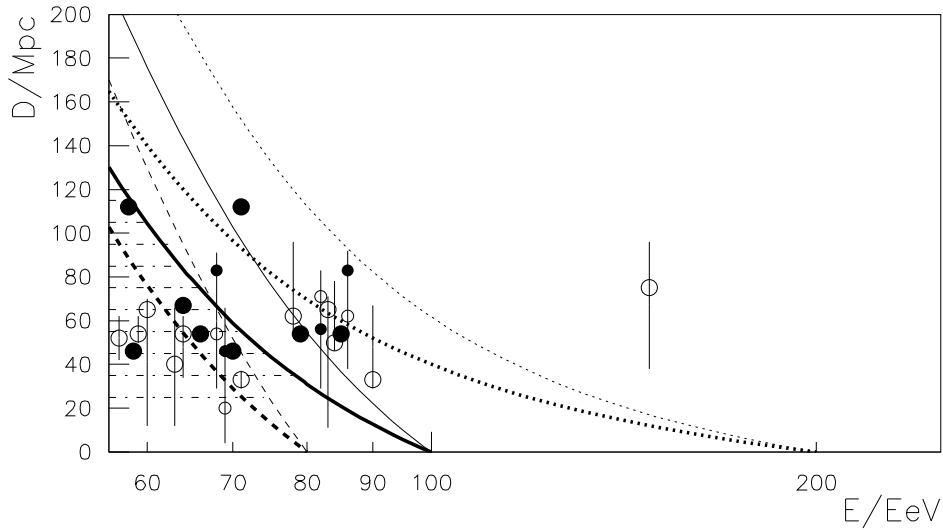,width=12.5cm}
\caption{Distance of travel of particles arriving with different energies.  
For the observations, the distances are derived from redshifts of AGNs or radio-galaxies close to the line of sight (see text).  
The curves refer to the expected energy of protons generated with particular initial energies
of 200, 100 or 80 EeV after they have travelled through the background radiation for different distances:
thin lines for energies actually assigned in the Auger analysis, thick lines apply if 
the assigned (and plotted) energies are too low by a factor 1.25.
The shaded area draws attention to the main area in which the points would be expected to lie in this latter 
case, assuming also that the source spectrum terminates near 100 EeV (``assigned energy'') }
\label{evd12}
\end{center}
\end{figure}

There is as yet insufficient information to resolve the interplay of several relevant factors, but the general 
constraints will be described with the aid of Figure \ref{evd12}.  
Consider first the lines, which trace the average change in a proton's energy with distance from its source.  
Ignoring the thicker lines, the steep dotted line, full line and dashed line refer to protons generated with 
energies of 200 EeV, 100 EeV and 80 EeV respectively, and the horizontal axis shows the energies on arrival if 
they have travelled for the distances specified on the vertical axis since leaving the accelerator. 
(He nuclei are not considered as they photodisintegrate in $\sim 1$ Mpc, and larger nuclei are ignored because, 
apart from the magnetic deflection argument given in section \ref{mag6sec}, their mean path length for serious 
mass loss by photodisintegration is much less than 100 Mpc.)   
One expects to have a mixture of energies leaving the accelerator, with a larger number starting at the lower 
energies.  It is seen that the expected distance distribution extends far above 100 Mpc, in contrast with the observations.

The only way to remove this tail of distant sources appears to be to assume that the reported energies are too low 
by a factor around 1.25, so that the true threshold energy is near 70 EeV.  
This makes the energy losses arising from photopion production reactions become much more rapid.  
The thicker dotted, full and dashed lines again refer to the proton energies after travelling various distances 
from the source, but assuming that the actual energies are always 1.25 times the figures written on the graph.  
The shaded area is roughly where we should expect to see the detected particles, on this basis, if there are few 
sources within 30 Mpc (as seems to be indicated by the data) and the source spectrum terminates not far from 
100 EeV nominal energy (or 125 EeV on a ``corrected'' energy scale): there would now be few particles having 
travelled more than 100 Mpc.  
This is closer to the observed distance distribution.

However, if we attempt to estimate the distance from the source of individual particles, by linking them to the 
AGNs seen close to the cosmic ray arrival direction, there is a problem.  
In outline, one expects that the particles arriving with the highest energies must have travelled a short 
distance, whilst the particles of lowest energy could come from a wide range of distances, with different starting 
energies.  
The observed data are presented as follows.  
The cosmic rays fairly clearly associated with extended radio galaxies (ERG) have a clear source distance, 
and are marked with large filled circles.  
For those not associated with extended radio galaxies, an open circle marks the median distance of VCV AGNs 
within $6^\circ$ and a vertical line shows the range of distances amongst this group of possible source AGNs. 
Where there is an adjacent extended radio galaxy and several VCV AGNs, the ERG distance and the VCV median 
distance are both shown, using smaller symbols (unless the ERG distance agrees with the VCV median, 
when the VCV indications are ignored). (There is no nearby known AGN for one of the 21 cosmic rays.)  
The expected correlation between energy and distance of travel is not apparent. 
The shape of the observed energy distribution is also a poor match, which might very possibly arise if there are 
event-to-event fluctuations in energy reconstruction accuracy of order 20\% or more, which largely scramble the 
energies in such a small span of energies as is represented here --- and
reminiscent of the Haverah Park example discussed in section \ref{rare4sec}.  
A larger data sample is needed to make more of this energy spectrum, to find out just how steeply it must fall at 
this apparent termination.  
Simulations of energy propagation have been made but will be published separately.

\section{Summary: conclusions concerning the cosmic-ray sources 
  \label{summ9sec}}

The report \cite{auger07} by the Auger collaboration, that most of the cosmic rays that had been detected above 
an energy of 57 EeV came from directions within $3.2 ^\circ$ of an optically detected AGN within about 75 Mpc, listed 
in the 12th catalogue of V\'{e}ron-Cetty and V\'{e}ron, marked the first instance of a fine-scale anisotropy seen in 
charged-particle astronomy.  
It was shown in section \ref{clustersec} that cosmic rays appeared to come from regions where AGNs were clustered.  
Sky distributions on different scales are consistent with the cosmic rays originating from sources in or near such AGNs, 
typically Seyfert galaxies, and being deflected randomly by $3-4^\circ$ on their way to us.  
As this scattering angle is similar to the separation of AGNs in the source region, 
it will not usually be clear what object is the actual source.
The appearance of some very weak objects closest to the observed direction is hence not significant.  
In section \ref{ueppsec}, a uniform-exposure polar plot was recommended for displaying quantitative aspects of 
the distribution of cosmic rays and possible source objects such as AGNs on the sky, and the distribution of 
cosmic rays was seen to follow the distribution of the VCV AGNs on the large scale, with about 0.1 cosmic 
rays detected during the reported observing run per catalogued AGN within 75 Mpc, where the exposure efficiency 
was high.

A method of showing the close correlation with directions of AGNs or other objects that did not require choice of 
a particular correlation ``window'' size was introduced in section \ref{rare4sec} --- a ``right ascension resonance'' in average closest distance ---
which showed the rapidity of decoherence of these patterns on the sky when they were displaced slightly.
This showed the cross-correlation in a very immediate way, that could be used to explore
 related associations in other regions.  
As this did not make use of any specially optimized window radius, it could be used to check
the very high statistical significance of the correlation with AGNs.  
The association of cosmic rays with the extended radio galaxies with radio jets within 120 Mpc 
(Nagar and Matulich \cite{nagmat}), most of which were not in the VCV optically-based catalogue, 
was, quite separately, significant at around the 1 in $10^6$ level, with an average of around
1 detected cosmic ray per object, although not all seem to be active at the present time: 
the nearby and spectacular low-energy electron accelerators Cen A and Fornax A are probably currently switched off at these ultra high energies (otherwise the former would have
completely outshone the other FRI radio galaxies, because of its proximity, as it
does in the case of low-energy synchrotron radiation).

If the few brightest radio galaxies within $\sim 80$ Mpc, notably M87 (Virgo A), Cen A and Fornax A, had been the 
dominant sources, their output must have been spread widely over the sky by magnetic 
deflections \cite{biermann3,biermann4,gorbunov}.  
However, only a small proportion of such widely-scattered cosmic rays would line up within $3.2^\circ$ of a 
catalogued AGN, as pointed out in section \ref{ueppsec}: this is not the case.  
If, on the other hand, the lines-of-sight to cosmic ray sources are scattered around many VCV AGNs by $3-4^\circ$, 
most of the cosmic ray distributions are well described. 
In summary, the small apparent scale of 3--$4^\circ$ for the spread of cosmic ray directions around source regions 
indicates that there are many more sources than there are strong radio-emitting AGNs.  
So although perhaps 40\% of the detected cosmic rays in this energy region appear to come from FRI or similar radio galaxies 
at typical distances near 50 Mpc, the remainder presumably come from numerous weaker sources, which reside in 
these AGN clusters, and accelerate protons to very similar maximum energies.  
Ghisellini et al \cite{ghis} have also emphasised the possibility that any objects that reside in galaxy clusters 
are candidates, and have shown that cosmic-ray directions correlate with the density of ordinary galaxies.  
It is not known whether the pattern de-coheres with galaxy density as rapidly as with AGNs, but it seems likely 
that AGNs do trace galaxy density.  
Although these authors particularly draw attention to the possibility that magnetars might be the sources \cite{arons},  
the very significant appearance of FRI radio galaxies close to the directions from which cosmic rays arrive 
makes it appear more probable that the Seyfert galaxies are indeed the currently less-active end of a distribution 
of jet-related sources of cosmic rays of $(0.7-1.2)\times 10^{20}$ eV.  
(Only 25\% of the VCV AGNs are not known radio emitters.)
As it is apparently the case that the particles, which, it is argued, are largely protons at the very highest energies, 
can rarely reach us from more than 100 Mpc away, it appears that the true energies must be around 25\% higher 
than currently estimated by the Auger methods, and that the accelerator output probably drops rather sharply in the 
region of $1.2 \times 10^{20} eV$.
It is, however, surprising that this assembly of stronger and weaker sources produces a total spectrum with a sharp cut-off.\\

\emph{Postscript}  Although further Auger observations have not yet been published,
talks (e.g. \cite{natureblog}) indicate that a later exposure has given a very different 
sky distribution, without the same concentration near AGNs.  
As was shown above, the published data show correlations with AGNs and related objects
that are not accidental; but the success of a 57 EeV threshold, for example, in extracting
the vital proton component that can point back to sources may not be robust,
as seen in the comparison of sky maps from three experiments at the end of section \ref{rare4sec}.
The success of the initial phase might even, for instance, have depended on the presence of
excessive fluctuations in (proton) energy reconstruction, occurring in somewhat
smaller fields of less well-behaved detectors, that falsely promoted the energies of
some protons!
It seems that the recognition of proton-induced showers will be an important task.
The appearance of very widely-scattered cosmic ray directions near the maximum energies
gives support to the indication from the Auger studies on depth of shower maximum 
\cite{augerxmax} that 
a heavy-nucleus component becomes very important as the extreme energies are approached.\\

\emph{Acknowledgements}\\

I wish to thank Ralph Engel for stimulating me to offer suggestions on the Auger directions,
 both Alan Watson and Johannes Knapp for several critical and perceptive discussions 
and much encouragement, information and help, 
Jeremy Lloyd Evans and Jim Hinton for useful suggestions, 
and Mansukh Patel and Ronald Bruijn for further help. 



\appendix
\begin{appendix}

\section{AGN counts in V\'{e}ron-Cetty V\'{e}ron 12th edition catalogue
  \label{Avvcsec}}

\begin{figure}[h!]
\begin{center}
\epsfig{file=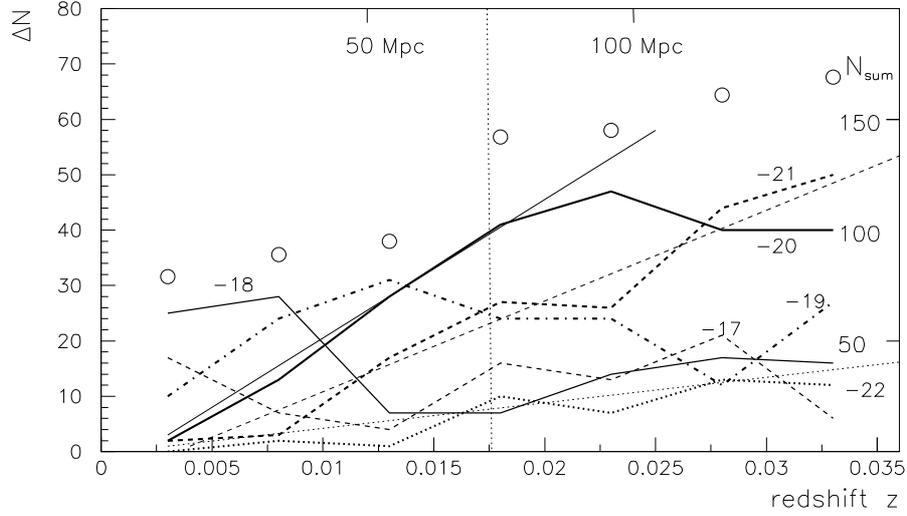,width=12.5cm}
\caption{Distribution in distance of AGNs (declination $< +15^\circ$) in VCV catalogue.
Lines marked -17 ... -21 refer to AGNs having  absolute magnitude -17.0 to -17.9,
-18.0 to -18.9...-21.0 to -21.9.  Line -22 refers to all magnitudes brighter than -22.0
(though only 27\% of those included are brighter than -22.9).  
For the three brighter groups, a dotted line shows the straight-line trend.  
Open circles show totals (scale on right).}
\label{vcvcharac}
\end{center}
\end{figure}

Figure \ref{vcvcharac} shows the number of objects listed in the southern hemisphere 
(declination $< +15^\circ$) per Mpc of distance, D, converting from redshift to distance 
using a value $H_o = 72\ $km s$^{-1}$ Mpc$^{-1}$ for Hubble's constant.  
The lines refer to objects in different ranges of absolute magnitude (``-19'' indicating -19.0 to -19.9, 
for example).  

The lines -20, -21 and -22 suggest a trend $dn/dD \propto D$ initially, as though the objects 
are distributed in a slab whose thickness is a few Mpc, but then falling sharply below this trend 
as though there is a limiting apparent magnitude for efficient detection.  
Thus objects with absolute magnitude brighter than -20.0 would be recorded inefficiently 
at distances beyond the 75 Mpc limit of the Auger sample.  
The median absolute magnitude for the most closely associated AGNs at the closer 
distances appears to be about 
-19.0, so beyond 75 Mpc, the catalogue would fail to list many of any truly 
associated AGNs (as remarked in \cite{auger07}, and one would need large 
statistics to map out the sources beyond there.  
If the number of listed AGNs within 75 Mpc had been much greater, the probability of 
associations within $3.2^\circ$ with any arbitrary direction would have been large, 
nullifying the signal.

\section{Array exposure to different celestial directions: equal-exposure plot radius.
  \label{A1sec}}

The effective detection area presented to a source at zenith angle $\theta$ is  $A\cdot\cos\theta$  if the detecting area 
on the ground always has a fixed value A.  
Let $\lambda$ be the magnitude of the observatory's latitude (i.e. always treated as positive).  
For a source at polar angle $p$, its zenith angle varies during the sidereal day according to the expression
\begin{equation} \cos \theta = \cos p \cdot\sin \lambda + \sin p \cdot\cos \lambda \cdot\cos H  
\label{sphertrig}
\end{equation}
where H is the hour angle relative to the source's right ascension, which increases uniformly with time, 
from 0 when $\theta$ is least (source highest in the sky), to $360^\circ$ after a complete sidereal day.  
Thus the time average of the effective detecting area is
\begin{equation}        \langle A \cdot \cos \theta \rangle = A \int \cos \theta \cdot dH / 360^\circ      
\label{avctheta}
\end{equation}
where the integral is taken over all values of H (here expressed in degrees) which give $\theta < \theta_{max}$ .  
For the Auger array, $\theta_{max} = 60^\circ$ and $\lambda = 35.2^\circ$ .  
At polar angles greater than $(90^\circ + \theta_{max} - \lambda )$ sources never rise to 
zenith angles less than $\theta_{max}$ : 
this limit is $114.8^\circ$ for the Auger observatory, meaning that sources with declinations above $+24.8^\circ$ 
are never recorded.  
Where $\theta_{max} > 90^\circ - \lambda$, as at Auger, some (``circumpolar'') sources never fall further from the zenith 
than $\theta_{max}$ : they have $p < \theta_{max} - (90^\circ - \lambda )$ and are detectable all day.  
If $\theta_{max} \le  90^\circ - \lambda$, though, sources closer to the pole than $p' = 90^\circ - \theta_{max} - \lambda$ 
are never recorded 
(and on an equal-exposure plot, this range $0 .. p'$ is contracted to a point at the centre of the plot).  
To find the effective detecting area for a source at any selected polar angle p, the integral in equation \ref{avctheta}
is evaluated over the range of H for which $\theta$ is less than $\theta_{max}$, noting that for some values 
of $p$ this condition is always or never satisfied, as just mentioned.

Having determined the efficiency of the detector for sources of each polar angle, as just described, 
the radius $R$ at which a source should be shown in an equal-exposure plot is 
\begin{equation}        R(p) = k \sqrt{\int_{z=0}^{z=p} \langle A \cos \theta \rangle \sin z\cdot dz}          
\label{rpa3},
\end{equation}
where the variable $z$ refers to a polar angle, and $\langle A \cos \theta \rangle$ is a
function of this polar angle, from equations \ref{avctheta} and \ref{sphertrig}.
The limit of integration, $p$, is the polar angle at which $R$ is required.
k is chosen to make $R \rightarrow 1$ at large $p$.
Table \ref{tab-4} tabulates the calculated efficiency and polar-plot radius, $R$, at intervals of $2^\circ$ in polar angle. The numerical results can be represented to a reasonable approximation by the following formula:

For the Pierre Auger Observatory,
\begin{equation}       R = [1+0.27 \exp(-p/5.7^\circ ) ]\cdot \sin(0.746p),    \qquad     (p \mbox{in degrees})  
\label{rapprox2}
\end{equation}
with the maximum divergence from the tabulated value being 0.006.  
However, if the term in square brackets is removed, leaving just the sine, the error is only significantly 
increased at polar angles $< 16^\circ$, reaching 0.012 at $8^\circ$, and as few cosmic rays are detected so close 
to the pole, this small distortion in their plotted density is probably immaterial.
This formula should not be used for polar angles beyond the theoretical maximum of $114.8^\circ$.

To plot a cosmic-ray direction, or an astronomical object, at right ascension $\alpha$ and declination $\delta$, 
the polar distance is calculated ( $p = 90^\circ -\delta$ for a northern hemisphere plot, 
or $p = 90^\circ +\delta$ for the southern hemisphere), then $R$ is found by interpolation in the table or by 
use of approximation given in equation (\ref{rapprox2}).  
The coordinates on the plot are then  $x = R \cdot\cos \alpha, \quad y = R \cdot\sin \alpha$.

\begin{table}[h]
\begin{center}
\begin{tabular}{ccc||ccc}
\hline
p  & Efficiency & R &  p & Efficiency & R \\[0.5ex]
\hline
     0  & 0.576 & 0.000 &         60  & 0.331 & 0.702  \\
     2  & 0.576 & 0.030 &        62  & 0.325 & 0.720  \\
     4  & 0.575 & 0.061 &        64  & 0.319 & 0.739  \\
     6  & 0.489 & 0.090 &        66  & 0.312 & 0.756  \\
     8  & 0.436 & 0.115 &        68  & 0.305 & 0.774  \\
    10  & 0.413 & 0.140 &        70  & 0.298 & 0.791  \\
    12  & 0.401 & 0.163 &        72  & 0.290 & 0.807  \\
    14  & 0.394 & 0.187 &        74  & 0.282 & 0.822  \\
    16  & 0.389 & 0.211 &        76  & 0.274 & 0.837  \\
    18  & 0.387 & 0.234 &        78  & 0.265 & 0.852  \\
    20  & 0.385 & 0.258 &        80  & 0.256 & 0.866  \\
    22  & 0.383 & 0.282 &        82  & 0.246 & 0.879  \\
    24  & 0.382 & 0.306 &        84  & 0.237 & 0.892  \\
    26  & 0.381 & 0.329 &        86  & 0.227 & 0.904  \\
    28  & 0.380 & 0.353 &        88  & 0.216 & 0.915  \\
    30  & 0.379 & 0.376 &        90  & 0.206 & 0.926  \\
    32  & 0.378 & 0.400 &        92  & 0.195 & 0.936  \\
    34  & 0.377 & 0.423 &        94  & 0.183 & 0.945  \\
    36  & 0.375 & 0.446 &        96  & 0.172 & 0.954  \\
    38  & 0.373 & 0.469 &        98  & 0.160 & 0.962  \\
    40  & 0.371 & 0.492 &       100  & 0.147 & 0.969  \\
    42  & 0.369 & 0.514 &       102  & 0.135 & 0.976  \\
    44  & 0.366 & 0.536 &       104  & 0.121 & 0.981  \\
    46  & 0.363 & 0.558 &       106  & 0.107 & 0.987  \\
    48  & 0.359 & 0.580 &       108  & 0.092 & 0.991  \\
    50  & 0.356 & 0.601 &       110  & 0.075 & 0.995  \\
    52  & 0.351 & 0.622 &       112  & 0.056 & 0.998  \\
    54  & 0.347 & 0.643 &       114  & 0.029 & 1.000  \\
    56  & 0.342 & 0.663 &       116  & 0.000 & 1.000  \\
    58  & 0.337 & 0.682 &            &       &        \\
\hline
\end{tabular}  
\end{center}  
\caption{Detection efficiency and radial position R in equal-exposure polar plot
for the Auger Observatory.}
\label{tab-4}
\end{table}  

\end{appendix}
\end{document}